\pdfoutput=1
\documentclass[amsmath,amssymb,aps,,pra,reprint,superscriptaddress]{revtex4-2} 

\usepackage[svgnames,psnames]{xcolor}
\usepackage{tabularx}
\usepackage{url}
\usepackage{braket}
\usepackage{graphicx}
\graphicspath{{.}{figs/}{../figs/}}
\usepackage{afterpage}
\usepackage{dcolumn}
\usepackage{bm}
\usepackage{tikz}
\usetikzlibrary{positioning,intersections,angles,quotes}
\usepackage{comment}
\usepackage{physics}
\usepackage{mathtools}
\usepackage[colorlinks,citecolor=DarkGreen,linkcolor=FireBrick,linktocpage,unicode,hypertexnames=false]{hyperref}
\usepackage{ulem}
\usepackage{mathcomp}
\usepackage{graphicx}

\usepackage{enumerate}

\begin{document}

\title{Coupled cluster method tailored with quantum computing}
\author{Luca Erhart}
\email{erhartl@hotmail.com}
\affiliation{%
Graduate School of Engineering Science, Osaka University, 1-3 Machikaneyama, Toyonaka, Osaka 560-8531, Japan
}%
\author{Yuichiro Yoshida}%
\email{yoshida.yuichiro.qiqb@osaka-u.ac.jp}
\affiliation{%
Center for Quantum Information and Quantum Biology,
Osaka University, 1-2 Machikaneyama, Toyonaka, Osaka 560-0043, Japan
}%

\author{Viktor Khinevich }%
\affiliation{%
Graduate School of Engineering Science, Osaka University, 1-3 Machikaneyama, Toyonaka, Osaka 560-8531, Japan
}%

\author{Wataru Mizukami}%
\email{mizukami.wataru.qiqb@osaka-u.ac.jp}
\affiliation{%
Graduate School of Engineering Science, Osaka University, 1-3 Machikaneyama, Toyonaka, Osaka 560-8531, Japan
}%
\affiliation{%
Center for Quantum Information and Quantum Biology,
Osaka University, 1-2 Machikaneyama, Toyonaka, Osaka 560-0043, Japan
}%

\date{\today}
	
\begin{abstract}
Introducing an active space approximation is inevitable for the quantum computations of chemical systems.
However, this approximation ignores the electron correlations related to non-active orbitals. 
Here, we propose a computational method for correcting quantum computing results using a well-established classical theory called coupled cluster theory. 
Our approach efficiently extracts the quantum state from a quantum device by computational basis tomography.
The extracted expansion coefficients of the quantum state are embedded into the coupled cluster ansatz within the framework of the tailored coupled cluster method. 
We demonstrate the performance of our method by verifying the potential energy curves of LiH, H\textsubscript{2}O, and N\textsubscript{2} with a correlation-energy correction scheme. Our method demonstrates reasonable potential energy curves even when the standard coupled cluster fails. 
The sufficient numbers of measurements for tomography were also investigated.
Furthermore, this method successfully estimated the activation energy of the Cope rearrangement reaction of 1,5-hexadiene together with perturbative triples correction. 
These demonstrations suggest that our approach has the potential for practical quantum chemical calculations using quantum computers.
\end{abstract}

\maketitle	

\section{Introduction}
Quantum chemistry is expected to be an application of quantum computing where it could be possible to outperform its classical counterpart \cite{Cao2019}. Quantum computers can hold and manipulate a superposition of an exponential number of electronic configurations using a polynomial number of quantum bits (qubits). They are, therefore, considered particularly suitable for simulations of strongly correlated systems, where the nature of quantum superpositions, often difficult to handle using current classical computers, is of essential importance.

Nevertheless, for the foreseeable future, quantum computers are limited in the number of qubits they have. In most cases, they cannot handle all the electronic degrees of freedom (orbitals and electrons) of a targeted molecule on a quantum computer. Furthermore, even if the number of qubits in a quantum computer increases sufficiently in the future, it is known that there will still be practical limits on the number of electrons and orbitals that a quantum computer can handle because of the slow clock speed of fault-tolerant quantum computers \cite{Babbush2021PRXQ, Goings2022PNAS}. Given these limitations, it is expected that quantum computers can only be applied to what is known as the `active space,' which is a user-defined, chemically important space around the Fermi level of a molecular electronic structure.

Active spaces are designed to consider only the most important orbitals for treating the strong correlation of electrons (i.e., static correlation). The active spaces, however, naturally ignore the remaining weak correlation (i.e., dynamical correlation) resulting from the non-active orbitals. Moreover, as the selection of an active space is more or less based on the chemical intuition of the user, there is often a degree of arbitrariness. This user dependency can have a significant impact on the accuracy of the methods using an active space. Recently automated active space selection procedures have been proposed to reduce user bias and improve the reliability of results~\cite{Stein2016,sayfutyarova2017automated,bao2018automatic,sayfutyarova2019constructing,jeong2020automation,king2021ranked,king2022large,kaufold2023automated}. The active space approximation has been extensively discussed and addressed in traditional quantum chemistry \cite{Lyakh2012,Yanai2015,Roos2016,Pathak2017,Park2020,Khedkar2021,Casanova2022}.

In quantum computing, various approaches have been proposed to incorporate ignored electron correlation~\cite{McArdle2020,Kumar2022,Sokolov2023, Bauman2019,Metcalf2020,NicholasP2021,Huang2022,Baumann2022,Le2023,Ma2021JCTC,Tammaro2023,Nishio2023PCCP,Takeshita2020,Mizukami2020PRR,Sokolov2019,Kottmann2021,Schleich2022,Rossmannek2021,Huggins2022Nature, YZhang2022arXiv, Wan2023ComMathPhys, Verma2021JCP,EXu2023arXiv}. These approaches fall into two broad categories. The first involves constructing models incorporating weak electron correlation during the development of effective Hamiltonians, which are then solved using quantum computers. This method is often called the `perturb-then-diagonalize' approach. The second category improves the results based on an active space Hamiltonian solution obtained by quantum computers and is known as the `diagonalize-then-perturb' approach. The former requires an up-front estimation of the electron correlation outside the active space, which introduces some arbitrariness. The latter commonly uses the internally-contracted multireference perturbation theory or multireference configuration interaction, and there have already been proposals for implementation using quantum computing results. However, these classical computing methods require higher-order reduced density matrices (RDMs), leading to prohibitive measurement costs on quantum computers and seem impractical \cite{Gonthier2022PRR,Tilly2021PRR,Nishio2023PCCP, takemori2023balancing}. Consequently, there is an increasing demand for methods considering weak electron correlation without relying on higher-order RDMs.

One of the practical and easy-to-use approaches to account for dynamical correlation from outside the active space without using higher-order RDMs is the tailored coupled cluster~(TCC)~\cite{Kinoshita2005,Hino2006,Lyakh2011CPL,Melnichuk2012JCP,Melnichuk2014JCP,Morchen2020JCP}. In TCC, after finding the ground state of an active space Hamiltonian by the complete active space configuration interaction (CASCI) method, the lost dynamical correlation in the CASCI calculation is described in an additional CCSD optimization of the whole space while keeping the static correction of CASCI solution. TCC can be worked with not only CASCI but also other quantum chemical theories such as DMRG~\cite{Faulstich2019}, FCIQMC~\cite{Vitale2020,Vitale2022}, and pair coupled cluster doubles \cite{Henderson2014JCP,Boguslawski2015linearized,Leszczyk2021JCTC,Nowak2021JCP}.
It should be noted that there is another method, known as the externally corrected coupled cluster method, which is similar to the TCC in its conceptual framework~\cite{Peris1997IJQC, XLi1997JCP, Peris1998MolPhys, Stolarczyk1994CPL, Xu2015JCP, Deustua2018JCP, Magoulas2021JCTC, SLee2021JCTC}.

This study proposes combining TCC and quantum computing to recover dynamical correlation from excluded orbitals outside the active space. Our practical method is a two-step approach using a quantum computer to solve the static correlation and add the remaining dynamical correlation using classical computers. 
This method requires only a limited number of qubits to mainly solve the static correlation, while the relatively computationally cheap consideration of weak correlation can be performed on classical hardware.
To extract the solved quantum state from a quantum computer, we use a tomographic method, computational basis tomography (CBT)~\cite{Kohda2022}. CBT is based on computational basis sampling~\cite{Kohda2022} and is an effective strategy when only a small number of Slater determinants are non-negligible, as is often the case in quantum chemistry. 
Additionally, a method is proposed, in which we utilize the direct access to the energy of the active space and add our newly determined electron correlation outside the active space. The accuracy of the proposed method is further enhanced using this approach. 

We demonstrate that our approach can produce accurate potential energy curves (PECs) for LiH, H\textsubscript{2}O, and N\textsubscript{2}. As CBT is naturally a statistical process that uses measurements to determine a quantum state, we investigate the number of measurement dependencies in CBT on LiH, H\textsubscript{2}O, and N\textsubscript{2}. Finally, to demonstrate the analysis of a realistic chemical reaction, we estimate the activation energy of the Cope rearrangement of 1,5-hexadiene.

The remainder of this paper is organized as follows: 
We discuss the relevant theories of TCC, CBT, and our new approach QC-CBT-TCC in Section~\ref{seq_theory}. The numerical results for the PECs of LiH, H\textsubscript{2}O, and N\textsubscript{2} are shown in Section~\ref{seq_results}, as well as the influence of different numbers of measurements in CBT have on the confidence of our approaches. Additionally, we determined the activation energy of the Cope rearrangement using our method and presented the results in Section~\ref{seq_results}. In Section~\ref{seq_conclustion}, we summarize our work.

\section{Theory} \label{seq_theory}

\subsection{Tailored Coupled Cluster} \label{subsection_TCC}

The TCC proposed by Kinoshita et al. \cite{Kinoshita2005} is a two-step theory designed to incorporate strong correlation into the standard coupled cluster theory, particularly the coupled cluster singles and doubles (CCSD) model. CCSD often breaks down when strong electron correlation exists. The main idea behind TCC is to split the coupled cluster (CC) parameters into those for strongly-correlated electrons in the active space and those for weakly-correlated electrons. They can then be determined individually. This gives rise to a formulation, as shown in the following equation.
\begin{equation} \label{equ_TCC_tailoring}
	\ket{\psi_{\text{TCC}}} = e^{\hat{T}^\text{{rest}} (\theta^\text{{rest}})} e^{\hat{T}^{\text{active}} (\theta^\text{{active}})} \ket{\psi_0}
\end{equation}
The operators $e^{\hat{T}^\text{{active}}}$ act exclusively on the active space. The other operators $e^{\hat{T}^\text{{rest}}}$ also act on the rest of the space. 
For CCSD, those operators are written as, 
\begin{equation}
	\begin{split}
		&\hat{T}^\text{{active}}(\theta) = \sum_{i,a} \hat{T}_{i}^a(\theta_i^a) + \sum_{i,j,a,b} \hat{T}_{i,j}^{a,b}(\theta_{i,j}^{a,b}) \\ &i,j,a,b \in \text{active space},
	\end{split}
\end{equation} 

\begin{equation}
	\label{equ_TCC_rest}
	\begin{split}
		&\hat{T}^\text{rest}(\theta) = \sum_{i,a} \hat{T}_{i}^a(\theta_i^a) + \sum_{i,j,ab} \hat{T}_{i,j}^{a,b}(\theta_{i,j}^{a,b}) \\ &\{i,j,a,b\} \not\subset \,\text{active space}, 
	\end{split}
\end{equation}
where $\hat{T}_{i}^a$ and $\hat{T}_{i,j}^{a,b}$ are the single and double excitation CC operators. Say, $\hat{T}_{i,j}^{a,b}$ excites electrons from the occupied orbitals $i$ and $j$ to the unoccupied orbitals $a,b$. The brackets \{\} in Eq. \eqref{equ_TCC_rest} indicate that at least one of the $i,j,a,b$ orbitals is not present in the active space.

In the first step of TCC, the molecule is described using a variational method using an active space approximation, such as the CASCI approach. This step aims to incorporate strong correlation by assuming that all the strongly correlated degrees of freedom are involved in the active space. Electron configurations' coefficients of the variational wavefunction $\ket{\psi^\text{active}_\text{{variational}}}$ can be mapped directly to coupled cluster amplitudes through the known relationship between their corresponding coupled cluster and configuration interaction (CI) operators: 
\begin{equation} \label{equ_T1}
	\hat{T}_1^\text{{active}} = \hat{C}_1,
\end{equation}
\begin{equation} \label{equ_T2}
	\hat{T}_2^\text{{active}} = \hat{C}_2 - \frac{1}{2} \hat{C}^2_1,
\end{equation}
where $\hat{C}_1$ and $\hat{C}_2$ are the CI operators to create single and double excitations, respectively.
This procedure allows us to approximately 
reconstruct the variational wavefunction$\ket{\psi^\text{active}_\text{{variational}}}$ 
in the CCSD ansatz.

The second step is the optimization of the remaining operators $e^{\hat{T}^\text{{rest}} (\theta^\text{{rest}})}$,
while the active space operators are kept fixed during the optimization. This preserved the description of static correlation. The optimized operator $e^{\hat{T}^\text{{rest}} (\theta^\text{{rest}})}$ incorporates the previously missing dynamic correlation into the solution.

Additionally, given that TCC is a method rooted in CCSD, after optimizing the CCSD operator $e^{\hat{T}^\text{{rest}} (\theta^\text{{rest}})}$, one can further improve accuracy by perturbatively incorporating the effects of the three-body excitation operator in the same way as in CCSD(T). So, we denote TCC with the (T) correction as TCC(T) in this study. For TCC(T), all active single and double amplitudes must be set to zero during the (T) correction calculation to prevent double-counting \cite{Lyakh2011CPL,Lang2020}. 

\subsection{Computational Basis Tomography}
CBT is an estimation method to determine a quantum state prepared on a quantum computer using the computational basis sampling method proposed by Kohda et al.~\cite{Kohda2022}. In this section, we briefly review how CBT estimates a quantum state $|\Psi \rangle$ on a quantum computer.

Let us call the estimated state CBT state $|\Psi_\textrm{CBT} \rangle$. Its definition is given as
\begin{equation}
	|\Psi_\textrm{CBT} \rangle = \frac{1}{\mathcal{N}} \sum_{i=1}^R \langle k_i | \Psi \rangle | k_i \rangle,
\end{equation}
where $1/\mathcal{N}$ is the normalization constant and $|k_i\rangle$ is a computational basis (i.e., Slater determinants or electron configuration) listed in the order of the absolute values of the coefficients $\langle k_i | \Psi \rangle$. The coefficients $\langle k_i | \Psi \rangle$ are equivalent to the CI coefficients. The parameter $R$ is introduced to truncate the trivial computational basis; therefore, $R$ needs to be chosen large enough in order to adequately approximate the original quantum state $| \Psi \rangle$. We tentatively consider the upper limit of $R$ to be around 1000-10000. By keeping the number of extracted coefficients $R$ constant, the method sacrifices some accuracy but enables scalability to large systems.

CBT aims to estimate the quantum state via measurements rather than the expectation value of an observable. The coefficients are expressed as follows:
\begin{equation}
	\langle k_i | \Psi \rangle = |\langle k_i | \Psi \rangle| e^{i\phi_i}.
\end{equation}
To get the coefficients, we must determine their absolute values and the phases $\phi_i$. The absolute value $|\langle k_i | \Psi \rangle|$ can be readily obtained using projective measurements; we show it in a later stage. However, a more detailed method is required to determine the phase. Instead of determining the phase directly, Kohda's computational basis sampling approach enables the efficient determination of CI coefficients through the observation of phase differences. We can use the following relationship to determine the phase difference between two computational bases $\ket{ k_i}$ and $\ket{ k_j}$:
\begin{equation} \label{Eq_relative_phase}
	e^{i(\phi_i - \phi_j)} = \frac{\langle k_i | \Psi \rangle \langle \Psi | k_j \rangle}{|\langle k_i | \Psi \rangle| |\langle \Psi | k_j \rangle|}.
\end{equation}
As the global phase of a state can be neglected, we can freely set the phase for one coefficient to zero and determine the relative phase differences for the remaining basis states using Eq.~(\ref{Eq_relative_phase}).
As described later, the absolute value of the coefficient $|\langle k_i | \Psi \rangle |$ and what we call the interference factor $\langle k_i | \Psi \rangle \langle \Psi | k_j \rangle$ can be determined relatively easily by projective measurements, thus allowing the estimation of phase differences.

When a sufficient number of samplings $N_\textrm{sample}$ are performed, the squared weight of $\langle k_i | \Psi \rangle $ is estimated as
\begin{equation} \label{equ_overlap}
	|\langle k_i | \Psi \rangle |^2 \simeq \frac{N_i}{N_\textrm{sample}},
\end{equation}
where $N_i$ is the number of times the outcome $k_i$ is obtained. This directly gives us an approximation for the absolute values of the coefficients.

To determine the phase factor, we need to measure the interference factors $\langle k_i | \Psi \rangle \langle \Psi | k_j \rangle$. In the case $k_i = k_j$ this is equivalent to a determine $|\langle k_i | \Psi \rangle |^2$. When $k_i \neq k_j$, we can rewrite the interference factor as
\begin{equation}
	\begin{split}
		\langle k_i | \Psi \rangle \langle \Psi | k_j \rangle &= |\langle 0 |U_{k_i,k_j}|\Psi \rangle|^2 + i|\langle 0 |V_{k_i,k_j}|\Psi \rangle|^2 \\
		&\quad- \frac{1+i}{2}(|\langle k_i | \Psi \rangle|^2 + |\langle k_j | \Psi \rangle|^2),
	\end{split}
\end{equation}
where $U_{k_i,k_j}$ and $V_{k_i,k_j}$ are unitary operators acting in the following way.
\begin{equation}
	U_{k_i,k_j} \left(\frac{\ket{k_i} +\ket{k_j}}{\sqrt{2}}\right) = \ket{0} , V_{k_i,k_j} \left(\frac{\ket{k_i} -\ket{k_j}}{\sqrt{2}}\right)= \ket{0}
\end{equation}

How to construct the circuits $U_{k_i,k_j}$ and $V_{k_i,k_j}$ was discussed in Ref.~\cite{Kohda2022,Eddins2022}.
Using the relation 
\begin{equation}
	\langle k_i | \Psi \rangle \langle \Psi | k_j \rangle = \frac{ (\langle k_1 | \Psi \rangle \langle \Psi | k_i \rangle)^* \langle k_1 | \Psi \rangle \langle \Psi | k_j \rangle }{|\langle k_1 | \Psi \rangle |^2},
\end{equation}
we note that we only need to measure the case for $k_i=k_1$ and $k_j \neq k_1$ and can recover the remaining interference factors. We explicitly write the cases $k_i=k_1$ and $k_j \neq k_1$ as:
\begin{equation} \label{0U1kPsi}
	\begin{split}
		\langle k_1 | \Psi \rangle \langle \Psi | k_j \rangle &= |\langle 0 |U_{k_1,k_j}|\Psi \rangle|^2 + i|\langle 0 |V_{k_1,k_j}|\Psi \rangle|^2 \\
		&\quad- \frac{1+i}{2}(|\langle k_1 | \Psi \rangle|^2 + |\langle k_j | \Psi \rangle|^2).
	\end{split}
\end{equation}
The first term in Eq.~(\ref{0U1kPsi}) is estimated as 
\begin{equation}
	|\langle 0 |U_{k_1,k_j}|\Psi \rangle|^2 \simeq \frac{N_0}{N_U},
\end{equation}
where $N_0$ is the count that we obtain the outcome zero and $N_U$ is the total amount of samplings.
Similarly, when $N_V$ is the number of measurements used to determine the second term in Eq.~(\ref{0U1kPsi}). The second term is estimated as
\begin{equation}
	|\langle 0 |V_{k_1,k_j}|\Psi \rangle|^2 \simeq \frac{N_0'}{N_V},
\end{equation}
where $N_0'$ is the count of the times we obtained the outcome zero.
The remaining terms of Eq.~(\ref{0U1kPsi}) can be determined similarly, as in Eq.~(\ref{equ_overlap}).

In summary, the truncation number $R$ and the numbers of the three types of measurements $N_\textrm{sample}$, $N_U$, and $N_V$ play vital roles in CBT.

\subsection{QC-CBT-TCC}

\begin{figure*}
	\includegraphics[width=\textwidth]{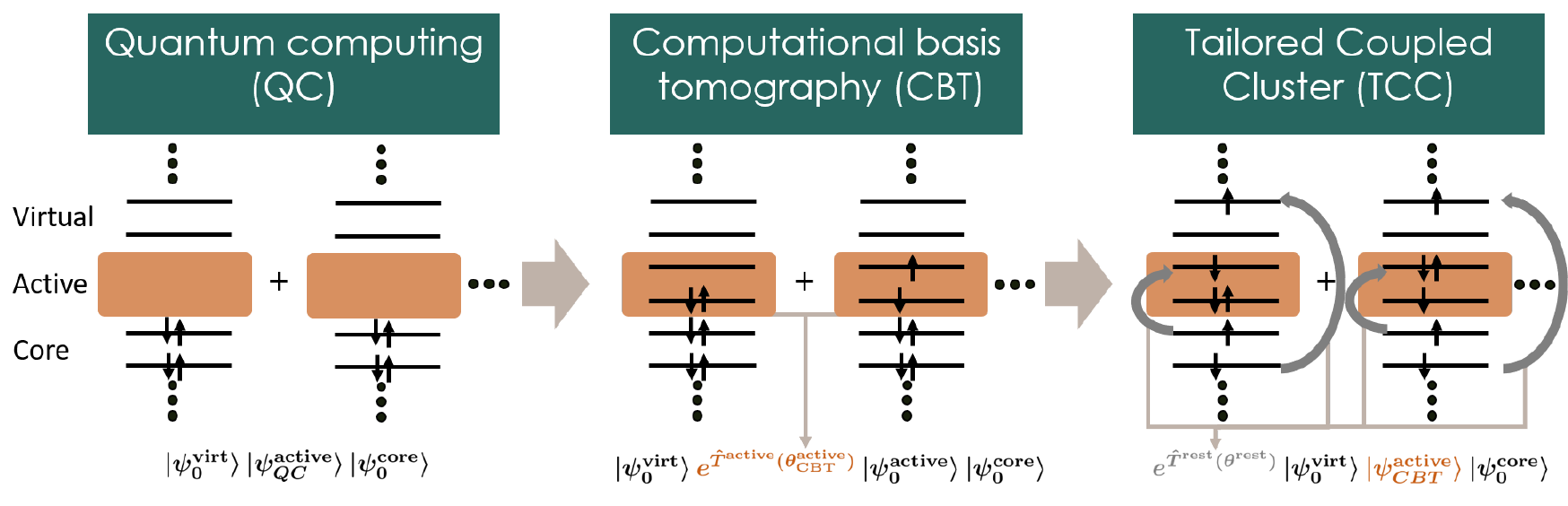}
	\caption{Schematic depiction of QC-CBT-TCC. In the first step, we determine the ground state solution of the active space using a quantum computer. As the state is classically inaccessible, we indicate this by not showing the concrete electronic structure of the active space. To determine the relevant CI-coefficients of the state, we use a CBT method and tailor the corresponding CCSD coefficients.
		In the last step, we optimize the remaining CCSD coefficients while keeping the active space coefficients constant.}
	\label{Figure_schee_VQE_TCC}
\end{figure*}

In this section, we discuss our qunatum-classical hybrid tailored coupled cluster theory with CBT method, which we denoted as QC-CBT-TCC. This method aims to include dynamic correlations for quantum computing while using an active space. QC-CBT-TCC uses a quantum computer to consider the strong correlation and then describes the remaining dynamic correlation using a classical device. In our framework, the CBT method is used to transfer the quantum solution onto a classical device. We present a graphical representation of our method in Figure \ref{Figure_schee_VQE_TCC} and explain it in more detail in the following.

In the first step, we determine the active-space ground state $\ket{\psi_{\text{QC}}^\text{active}}$ using a quantum computer.
Different quantum (or quantum-classical hybrid) algorithms, such as variational quantum eigensolver~(VQE)~\cite{Peruzzo2014,Fedorov2022}, quantum phase estimation~(QPE)~\cite{Aspuru-Guzik2005}, and qunatum imaginary time evolution~\cite{McArdle2019,Motta2019,Gomes2020JCTC,Gomes2021AQT,Tsuchimochi2023JCTC} can be used in this step.

In the second step, CBT is used to determine the most significant CI coefficients of the produced ground state on a quantum computer $\ket{\psi^\text{active}_\text{QC}}$. Using the obtained CI coefficients, the ground state of the active space $\ket{\psi^\text{active}_\text{QC}}$ is approximately converted to a coupled cluster ansatz. Mapping the CI operators to their coupled cluster counterparts follows the same relationship as already described in Section \ref{subsection_TCC} in equation \eqref{equ_T1} to \eqref{equ_T2}. 
As the wavefunction parmameters $\theta^\text{active}$ for $\ket{\psi^\text{active}_\text{QC}}$ are practically approximated by CBT, we call them $\theta_{\text{CBT}}^\text{active}$. The obtained approximate active-space ground state wavefunction in the coupled cluster is written as: 
\begin{equation}
	\ket{\psi_{\textrm{CBT}}^\text{active}} = e^{\hat{T}^\text{active} (\theta_{\textrm{CBT}}^\text{active})}\ket{\psi_{0}}.
\end{equation}

In the third step, the remaining $\theta^\text{rest}$ coefficients are optimized by solving the standard projected coupled cluster equations. During the optimization, the amplitudes of the active space operators are held constant. This procedure allows the addition of dynamical correlation descriptions while conserving that of static correlation. The final produced state of QC-CBT-TCC $\ket{\psi_\text{QC-CBT-TCC}}$ can be written as:

\begin{equation} \label{equ_VQE_TCC}
	\begin{split}
		\ket{\psi_\text{QC-CBT-TCC}}
		= 
		&e^{\hat{T}^\text{rest} (\theta^\text{rest})} e^{\hat{T}^\text{active} (\theta_{\text{CBT}}^\text{active})} \ket{\psi_0} \\
		= &e^{\hat{T}^\text{rest} (\theta^\text{rest})}  \ket{\psi_{\text{CBT}}^\text{active}}.
	\end{split}
\end{equation}
The QC-CBT-TCC energy $ E_{\textrm{QC-CBT-TCC}}$ is then given by the following projected energy:
\begin{equation}
	\begin{split}
		E_{\textrm{QC-CBT-TCC}} = \bra{\psi_0} \hat{H} \ket{\psi_\text{QC-CBT-TCC}}.
	\end{split}
\end{equation}

\subsection{Enhanced QC-CBT-TCC}
One of the issues associated with QC-CBT-TCC is the lack of assurance that the energy obtained via QC-CBT-TCC will outperform classical algorithms in accuracy. This arises because of the approximate nature of both CBT and TCC methodologies. The TCC approach employs the standard CCSD ansatz to approximate the active space wavefunction, which implies that it does not incorporate information about higher-order excitations, such as triple and quadruple excitations, obtained from a quantum computer. Furthermore, CBT is a statistical method, and its inherent statistical nature introduces additional errors. To address this problem, we introduce a correction method as described below.
The corrected energy $E_{\rm QC-CBT-TCC(c)}$ is defined as:
\begin{equation} \label{equ_corrected_ernergy}
	\begin{split}
		E_{\text{QC-CBT-TCC(c)}} &= E^{\text{active}}_{\rm \text{QC}} \\
		&+ (E_{\textrm{QC-CBT-TCC}} - E_{\textrm{QC-CBT-TCC}}^{\rm \text{active}}).
	\end{split}
\end{equation}
The first term $E^{\text{active}}_{\rm \text{QC}}$ in Eq.~\eqref{equ_corrected_ernergy} represents the expectation value of the active space Hamiltonian via quantum computing such as QPE. The difference in the second term corresponds to the additional correlation added to the active space solution using the QC-CBT-TCC technique. 
$E_{\textrm{QC-CBT-TCC}}$ is the predicted energy using the QC-CBT-TCC method, whereas
\begin{equation}
	\begin{split}
		E_{\textrm{QC-CBT-TCC}}^{\text{active}} = \bra{\psi_0} \hat{H} e^{\hat{T}^\text{active} (\theta_{\textrm{CBT}}^\text{active})}\ket{\psi_{0}},
	\end{split}
\end{equation}
is the CCSD energy after dressing the CCSD amplitudes but before the remaining CCSD optimization. In that case, the remaining operators $e^{\hat{T}^\text{rest}}$ are equivalent to the identity. Since the same error arise in $E_{\textrm{QC-CBT-TCC}}$ and $E_{\textrm{QC-CBT-TCC}}^{\rm \text{active}}$,  taking the difference between these terms removes the errors in the active-space electron correlation inherent to the QC-CBT-TCC method.

Note that Izs{\'{a}}k et al. have already employed a similar extrapolative correction method in Ref.~\cite{Izsak2023} in the context of using quantum computers for quantum chemistry as well as Daniel Kats and his co-workers for FCIQMC-TCC~\cite{Vitale2020}.

\section{Results and Discussion} \label{seq_results}
This section describes our new method, QC-CBT-TCC, and the QC-CBT-TCC(c) approach's performance. Sec. \ref{Section_PEC} presents the potential energy curves for three molecules, LiH, H\textsubscript{2}O, and N\textsubscript{2}. In Sec. \ref{Section_num_of_shots}, we conducted multiple experiments at different interatomic distances for each molecule to discuss the reliability of CBT, as CBT is a statistical method and QC-CBT-TCC is influenced by the statistical errors of CBT. In Sec. \ref{Section_Cope_rearrangement}, we apply our method to a prototype organic reaction, the Cope rearrangement of 1,5-hexadiene.

The computational details are as follows.
A chemqulacs~\cite{chemqulacs} library was used to simulate quantum circuits on classical hardware. We employed disentangled unitary coupled cluster singles and doubles (UCCSD) as an ansatz for VQE and BFGS as an optimizer. CCSD, CCSD(T), CASCI, and FCI were calculated using PySCF~\cite{Sun2020}. The MR-CISD+Q method was employed to obtain reference data for the PECs of H\textsubscript{2}O and N\textsubscript{2} because the FCI calculation was too computationally demanding for the hardware at hand. This method was implemented in the ORCA quantum chemistry program Version 5.0.3~\cite{Neese2022}. 
All calculations for the PECs in this study were performed using the cc-pVDZ basis set. In Section \ref{Section_Cope_rearrangement}, we used the 6-31G$*$ basis set to compute the activation energy for the Cope rearrangement. We used the canonical orbitals for all molecules except for LiH, which was calculated using the natural orbitals provided by a CCSD calculation. This, however, is an approximation because the CCSD implementation of PySCF used in this study assumes canonical orbitals, so off-diagonal terms in the Fock matrix are ignored. The active spaces were determined using the lowest energy level unoccupied orbitals and the highest energy occupied orbitals from the Hartree-Fock calculation. To transfer the CI coefficients determined by CBT to the amplitudes for CCSD, only the real parts were considered. This was necessary because usually the CCSD for a non-relativistic or non-periodic Hamiltonian expects real coefficients. The CI coefficients were normalized to ensure the normalization. Unless otherwise indicated, we applied the parameters $R=100$, $N_{\text{sample}} =10^6$, $ N_{U}=10^6$, and $N_{V}=10^6$ for all the CBT calculations.

\subsection{Potential energy curves \label{Section_PEC}}
First, we investigated the PEC of the LiH molecule. In addition to the QC-CBT-TCC(c), we show the results for QC-CBT-TCC, active space UCCSD, and FCI, as well as HF, CCSD, and CCSD(T). The selected active spaces for these calculations were two electrons and two spatial orbitals, the highest occupied and the lowest unoccupied orbitals.

\begin{figure*}
	\includegraphics[bb=0 0 1045 342, width=0.95\textwidth]{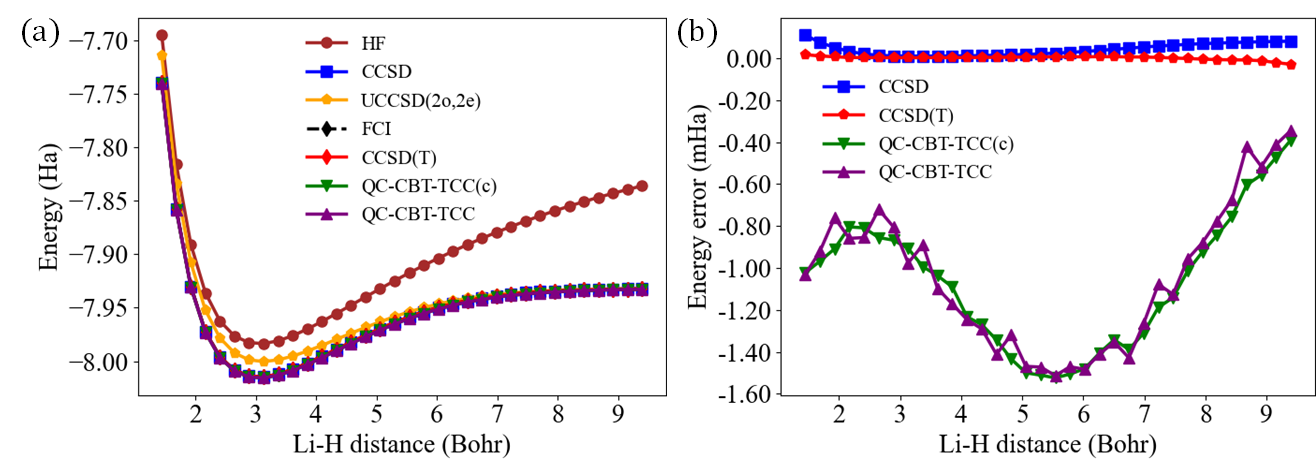}
	\caption{ (a) Potential energy curves of LiH using the cc-pVDZ basis sets. The CCSD natural orbitals were employed for the VQE and subsequent QC-CBT-TCC and QC-CBT-TCC(c) calculations. An active space consisting of 2 orbitals and 2 electrons [i.e., (2o, 2e).] was used. (b) Deviations from the FCI potential energy curve.}
	\label{Figure_LiH_pec}
\end{figure*}

Figure~\ref{Figure_LiH_pec} (a) shows the PEC of the LiH molecule. All considered methods followed the qualitative features of the FCI energy curve shape, except for HF. The active space UCCSD increased the accuracy compared to that of the HF solution. This behavior is most notable in the bond dissociation region. At the equilibrium bond length, the active space UCCSD can capture some dynamic  correlation. The QC-CBT-TCC and its enhanced version, the QC-CBT-TCC(c), further increased the computed energy's accuracy. This indicates that additional dynamical correlations outside the active space are required to reach good ground-state energy. CCSD and CCSD(T) both gave energies very similar to FCI. In the high dissociation regime, where static correlation dominates, all methods except for HF gave similar results.

We focus on the energy errors of the coupled cluster and present methods compared with the FCI energy in Figure~\ref{Figure_LiH_pec} (b). CCSD and CCSD(T) showed better accuracy throughout the tested bond lengths than the QC-CBT-TCC and QC-CBT-TCC(c). This can be attributed to the small static correlation within the molecule. For such a system, CCSD and CCSD(T) can capture electron correlation up to high accuracy. The energy difference between the QC-CBT-TCC and QC-CBT-TCC(c) was small. However, the QC-CBT-TCC(c) appears to produce a smoother energy error. We assume that some errors in the QC-CBT-TCC did cancel out when the QC-CBT-TCC(c) was constructed. QC-CBT-TCC and QC-CBT-TCC(c) produced energies that are lower than the FCI energy. This is possible since QC-CBT-TCC tailors the classical coupled cluster equations which are not variational.

Secondly, we observed the two-bond dissociation behavior of H\textsubscript{2}O. We simultaneously increased the two H-O distances while leaving the HOH angle constant at 104.52$\tcdegree$. Such a treatment can be understood as an example of a doubly bond dissociation~\cite{Kinoshita2005}. We selected an active space of eight electrons and six spacial orbitals of the four highest occupied and two lowest unoccupied orbitals. The MR-CISD+Q calculation was based on preliminary state-specific CASSCF(6o,8e). 

\begin{figure*}
	\includegraphics[bb=0 0 1045 342, width=0.95\textwidth]{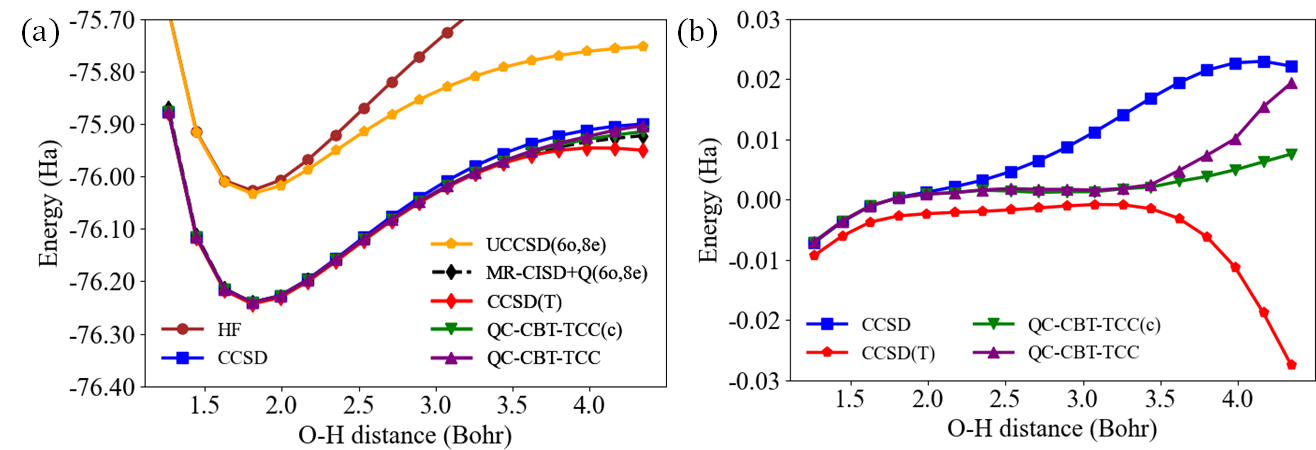}
	\caption{(a) Potential energy curves of the double dissociation of the water molecule's OH bonds using the cc-pVDZ basis sets. An active space consisting of 6 orbitals and 8 electrons [i.e., (6o, 8e)] was used. (b) Deviations from the MR-CISD+Q potential energy curve of the double dissociation of the water molecule's OH bonds.}
	\label{Figure_H2O_pec}
\end{figure*}

As shown in Figure \ref{Figure_H2O_pec} (a), QC-CBT-TCC and QC-CBT-TCC(c) reproduce the quantitative features of the MR-CISD+Q calculation and show good accuracy throughout the observed bond distances. The CCSD energy curve, on the other hand, shows less accurate results but still follows the MR-CISD+Q energy curve. This contrasts with CCSD(T), which fails to compute the energy in the high dissociation region. At the equilibrium point, the active space UCCSD(6o,8e) and HF estimated similar energies. For long bond distances, UCCSD gave more accurate energies than HF. Treating the molecule within natural orbitals would probably increase the accuracy of the UCCSD(6o,8e). Methods that included out-of-active space correlation produced significantly more accurate results than the UCCSD(6o,8e). This shows that such correlation must be included to obtain an accurate ground state description.

In Figure \ref{Figure_H2O_pec} (b), we show the explicit error of the presented methods compared to MR-CISD+Q. CCSD(T), and arguably CCSD, also break in the highly entangled region. QC-CBT-TCC and the QC-CBT-TCC(c) can give accurate energies. However, in the high dissociation region, QC-CBT-TCC's error increases, whereas that of QC-CBT-TCC(c) shows this effect is reduced, indicating that the QC-CBT-TCC(c) method produces more stable energies.

Finally, we investigated the PEC of N\textsubscript{2} as an example of a triple-bonded molecule. We used the three highest occupied and three lowest unoccupied orbitals for the active space to perform the active space UCCSD calculations. The MR-CISD+Q calculations were performed following the state-specific CASSCF(10o,10e). 

\begin{figure*}
	\includegraphics[bb=0 0 1045 342, width=0.95\textwidth]{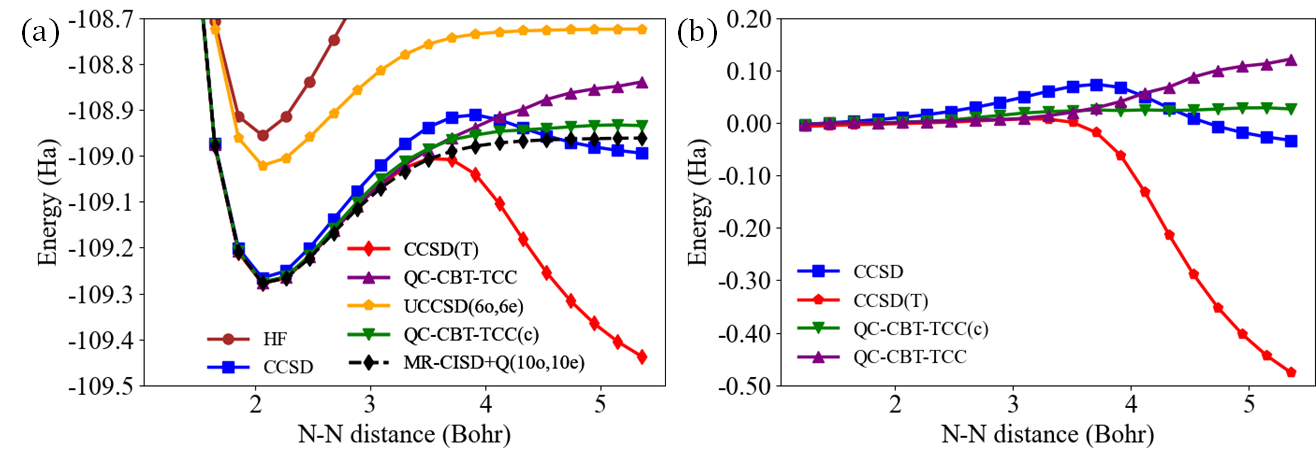}
	\caption{(a) Potential energy curves of N\textsubscript{2} using the cc-pVDZ basis sets. UCCSD calculations employed the active space consisting of 6 orbitals and 6 electrons [i.e., (6o, 6e)], while MR-CISD+Q calculations used CASSCF with the active space of 10 orbitals and 10 electrons as a reference wavefunction. (b) Deviations from the MR-CISD+Q potential energy curve of N\textsubscript{2}.}
	\label{Figure_N2_pec}
\end{figure*}

Figure \ref{Figure_N2_pec} (a) shows that the QC-CBT-TCC(c) reproduces the MR-CISD+Q PEC well. However, there is an almost constant, small energy gap between the two PECs in the region of approximately more than 4.0~Bohr. We expect this gap to result from the different active spaces selected for the QC-CBT-TCC(c) and MR-CISD+Q calculations. CCSD and CCSD(T) failed to reproduce the triple-bond breaking for extensive bond lengths as well known. They exhibited unnatural behavior and overestimated the correlation energy. The energy of active space UCCSD increased the accuracy compared with the HF energy because it could consider electron correlation inside the active space. Like the previous results, the additional dynamical correlation from non-active space affected the QC-CBT-TCC energy, and QC-CBT-TCC(c) further demonstrates the significant accuracy. 

The error of the observed methods compared with that of MR-CISD+Q is shown in Figure \ref{Figure_N2_pec} (b). Only the QC-CBT-TCC(c) provided a good solution for all observed distances. As well known, CCSD and CCSD(T) fail to predict accurate energies for N\textsubscript{2} for considerable bond lengths, but also QC-CBT-TCC started to produce unfaithful results in the high dissociation limit. This effect suggests that the QC-CBT-TCC(c) can cancel out some errors resulting from the transition from the quantum device to the CCSD ansatz and is a genuine improvement of the method.

\subsection{Number of shot dependency of QC-CBT-TCC \label{Section_num_of_shots}} 
CBT used in our method adds a certain level of uncertainty to our computed energies. Determining the CASCI coefficients using CBT is a statistical process. To investigate this statistical behavior, we examined for the molecules LiH, H\textsubscript{2}O, and N\textsubscript{2} the impact of varying measurement repetitions in CBT on the QC-CBT-TCC(c). We considered two constellations, one for the molecules at the equilibrium bond length \cite{Kinoshita2005} and one for a bond length of twice the equilibrium bond distance. Note that the geometries were not optimized. To ensure statistical significance, we repeated each measurement setup 1000 times and showed its effect on the QC-CBT-TCC(c). The parameter $R$, used to truncate the trivial computational basis in CBT, is set as 100. The behavior of LiH was similar to that of H\textsubscript{2}O and N\textsubscript{2}, and the results are shown in the Appendix.
We demonstrated the effect using a box blot to demonstrate numerical data graphically.

\begin{figure*}
	\includegraphics[bb=0 0 989 327, width=0.95\textwidth]{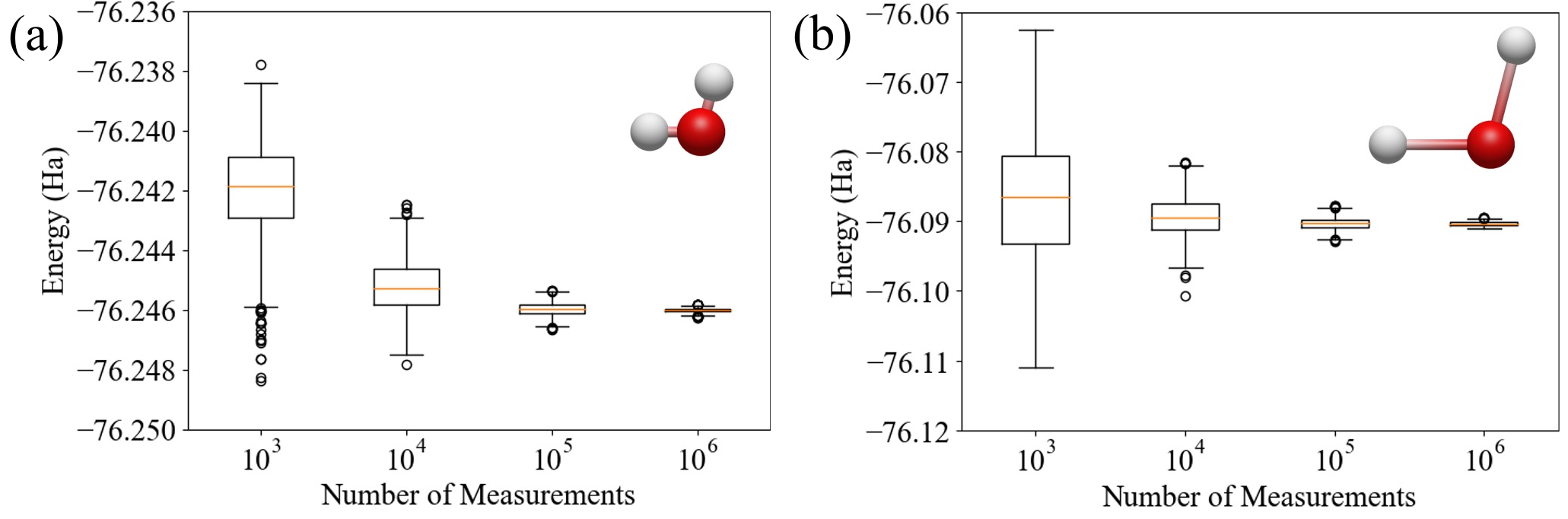}
	\caption{Energy insecurity of QC-CBT-TCC(c) for H\textsubscript{2}O with a different number of measurements for CBT. $N_{\text{sample}}, N_{U}$, and $N_{V}$ were all measured in the specified quantities in the figure as the X-axis values. CBT was performed 1000 times for each setting to obtain an estimate for the statistical errors. The box indicates the range from the first to the third quartile called the inter-quartile range (IQR), with the median drawn in the middle. The box contains $50 \% $ of the data. Whiskers are the lines extending 1.5 times the IQR from the first and third quantiles. Data points that exceed the whiskers are considered outliers (fliers) and are represented as single dots. (a) Number of measurements dependency at an equilibrium distance of 1.808 Bohr. (b) Number of measurements dependency at 3.617 Bohr, twice the equilibrium distance. }
	\label{Figure_H2O_energy_uncertainty_equilibrium}
\end{figure*}

Figure \ref{Figure_H2O_energy_uncertainty_equilibrium} (a) and Figure \ref{Figure_H2O_energy_uncertainty_equilibrium} (b) discuss the influence of the number of CBT measurements for QC-CBT-TCC(c) for H\textsubscript{2}O. The two graphs show the differences in the bond lengths between oxygen and hydrogen. The boxes become smaller in both cases. Thus, the uncertainty decreases with an increased number of measurement repetitions. This effect is more prevalent in Figure \ref{Figure_H2O_energy_uncertainty_equilibrium} (b), which was expected as the electronic structure of the ground state in the dissociation region is more complex and involves more relevant electron configurations than those around the equilibrium point. The greater complexity in the dissociation region requires more measurements to determine all relevant CI coefficients accurately. Statistical errors are smaller at the equilibrium point because the Hartree-Fock state, a single computational basis state, is already a good approximation.

\begin{figure*}
	\includegraphics[bb=0 0 989 327, width=0.95\textwidth]{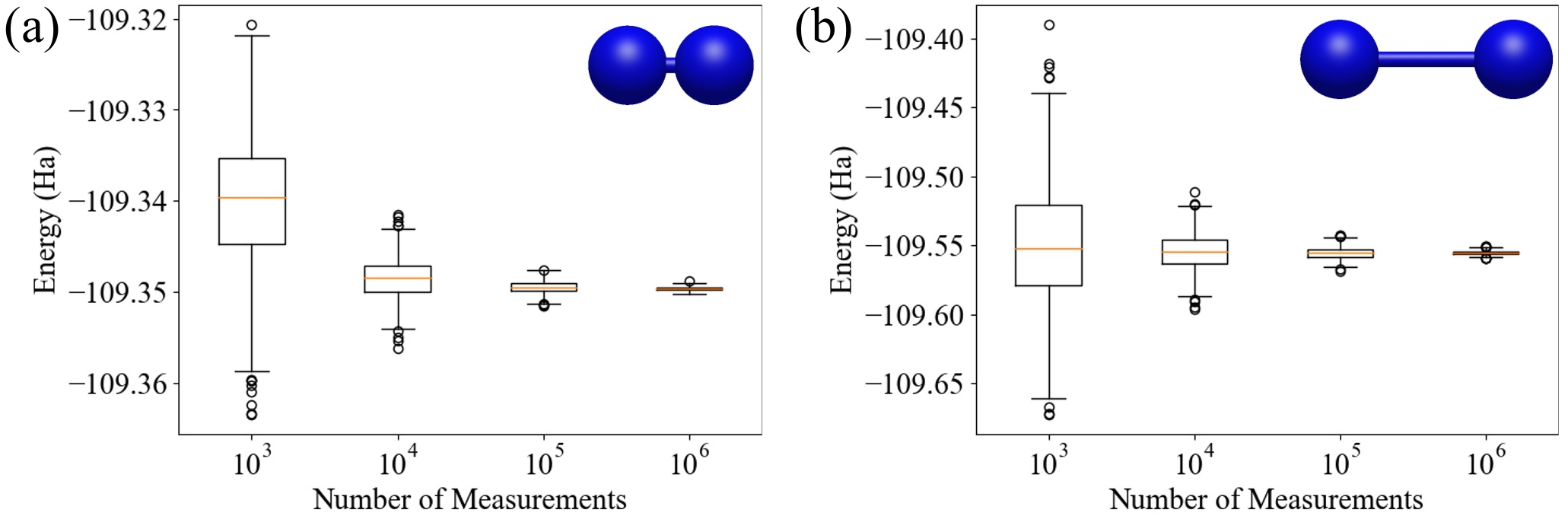}
	\caption{Energy insecurity of QC-CBT-TCC(c) for N\textsubscript{2} with a different number of measurements for CBT. $N_{\text{sample}}, N_{U}$, and $N_{V}$ were all measured in the specified quantities in the figure as the X-axis values. CBT was performed 1000 times for each setting to obtain an estimate for the statistical errors. The box indicates the range from the first to the third quartile called the inter-quartile range (IQR), with the median drawn in the middle. The box contains $50 \% $ of the data. Whiskers are the lines extending 1.5 times the IQR from the first and third quantiles. Data points that exceed the whiskers are considered outliers (fliers) and are represented as single dots. (a) Number of measurements dependency at an equilibrium distance of 2.060 Bohr. (b) Number of measurements dependency at 4.119 Bohr, twice the equilibrium distance.}
	\label{Figure_N2_energy_uncertainty_equilibrium}
\end{figure*}

The results for N\textsubscript{2} show a behavior similar to that of H\textsubscript{2}O in that increasing the number of measurements reduces the QC-CBT-TCC(c) energy distribution, as seen in Figures \ref{Figure_N2_energy_uncertainty_equilibrium} (a) and \ref{Figure_N2_energy_uncertainty_equilibrium} (b). This behavior can be explained by the higher accuracy and decreased uncertainty in determining the CBT state coefficients. Consequently, the produced state is more consistent; therefore, the energy expectancy can be determined with a higher degree of confidence.

These observations show the importance of accounting for the complexity of electronic structures when applying CBT. This becomes increasingly important in the high dissociation region when the states become more entangled. A large number of measurements are necessary to accurately determine all the relevant coefficients. 

To estimate the number of shots needed to reach sub-mH standard deviation of the predicted energy, we performed further calculations for the  N\textsubscript{2} at 5.355 Bohr. This system is the most challenging for CBT in our study. The other molecules need considerably fewer measurements. A standard derivation of $0.61$ mH using QC-CBT-TCC(c) can be reached using $3 \times$ $10^7 (N_{\text{sample}} = 10^7, N_U = 10^7,$
and $N_V = 10^7)$ shots. This is in reach for today's quantum hardware and similar to the predicted shot counts reported by Scheurer et al.~\cite{Scheurer2023} using the matchgate classical shadows to extract CI coefficients from a quantum computer.

\subsection{Application to Cope rearrangement \label{Section_Cope_rearrangement}}

Our dynamical correlation correction allows the analysis of realistic chemical processes using quantum computers. We estimated the activation energy of the Cope rearrangement using our method.
Cope rearrangements are well-known organic chemical reactions. In this reaction, the carbon chain of 1,5-hexadiene is rearranged in a concerted way to convert into itself via the transition state of the chair form.

\begin{figure}
	\includegraphics[bb=0 0 663 284, width=0.45\textwidth]{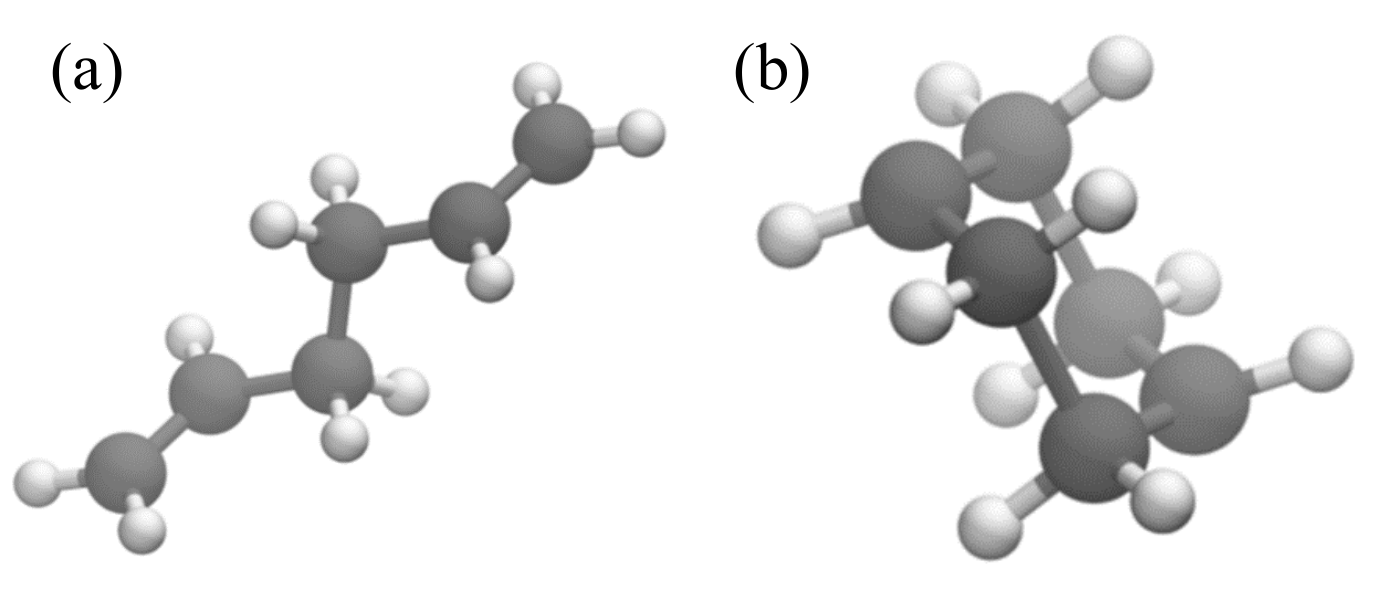}
	\caption{(a) 1,5-hexadiene in the $C_i$ symmetry structure. (b) Transition state of Cope rearrangement of 1,5-hexadiene in the chair-form.}
	\label{fig:1.5_hexadine}
\end{figure}

We considered the energy difference between the 1,5-hexadiene and its chair-form transition state, which are $C_i$ symmetry structures. Their geometric structures are shown in Figure~\ref{fig:1.5_hexadine} (a) and~\ref{fig:1.5_hexadine} (b). We performed QC-CBT-TCC(c) with the perturbative tripled corrections, denoted as QC-CBT-TCC(T)(c), to evaluate the activation energy.

For the chair and hexadiene ($C_i$ symmetry) geometries, we used the optimized structures from Ref.~\cite{Ventura2003}; they used the analytic MR-CISD and multireference averaged quadratic coupled-cluster (MR-AQCC) gradient methods~\cite{Shepard1991} to optimize the structures with a reference space of CAS(6o,6e) in a 6-31G$*$ basis set. 

The result is shown in Table~\ref{Tab_cope_rearranment}.
Our newly proposed method agrees well with the experimental value in Ref.~\cite{Staroverov2001} and the value of MR-AQCC~\cite{Ventura2003}. We observed that methods that ignore outside-of-active space correlation, such as HF, CASCI, and the active space UCCSD, failed to compute a reasonable activation energy. Including the dynamical correlation significantly improved the accuracy of these methods. The QC-CBT-TCC(T)(c) produced comparable results to the MR-AQCC method. This result demonstrates our method's potential for simulating complex chemical processes with strong correlation.

\begin{table}[ht] 
	\centering
	\caption{Activation energies in kcal/mol for 1,5-hexadiene Cope rearrangement calculated in the 6-31G$*$ basis set. The VQE calculations used the disentangled UCCSD ansatz and an active space consisting of 6 orbitals and 6 electrons [i.e., (6o, 6e)]. The experimental values were obtained from the computational study in Ref.~\cite{Staroverov2001}, which was based on the experimentally measured enthalpy from Ref.~\cite{Von1971} \label{Tab_cope_rearranment}}
	\resizebox{\linewidth}{!}{
		\begin{tabular}{cc}
			\hline\hline
			Methods                        &  Activation energy \\ \hline
			HF                         & 66.0 \\
			CASCI(6o, 6e)                    & 59.0 \\
			UCCSD(6o, 6e)           & 60.1 \\
			QC-CBT-TCC(T)(c) & 38.7 \\
			MR-AQCC~\cite{Ventura2003}           & 37.3 \\
			Experiment~\cite{Staroverov2001}                & 35.0\\
			\hline\hline
	\end{tabular}}
\end{table}

\section{Conclusions} \label{seq_conclustion}

This study combines a tailored coupled cluster approach with quantum computing. The tailored coupled cluster approach separates the active space from the remaining orbitals. This allows for a more rigorous active space treatment while maintaining the dynamic correlation from the out-of-active space orbitals. Computing an eigenvalue of an active space Hamiltonian using a quantum computer is desirable because it is expected that they can simulate much larger quantum systems. We tailor the CCSD approach using the quantum state of the active space determined on a quantum computer. We use this approach to describe dynamical correlation from the out-of-active space orbitals on the quantum solution. We employ computational basis state tomography (CBT) to determine the relevant CCSD amplitudes using a quantum computer. This makes our method universally usable for all quantum algorithms that produce an eigenstate of an active-space Hamiltonian. 

The QC-CBT-TCC was applied to three small molecules, LiH, H\textsubscript{2}O, and N\textsubscript{2}. In the LiH case, where static correction is less important, our method may improve the accuracy compared to the active space UCCSD, and we observe that our method with correction scheme, QC-CBT-TCC(c), works well to cancel out the error of QC-CBT-TCC. In the H\textsubscript{2}O and N\textsubscript{2} cases, our method improved the accuracy and could provide suitable quantitative PECs, even when standard CCSD or CCSD(T) fails. 
Hence, our method has the potential to practically include the lacking dynamical correlation into a static correlated active space solution of quantum computation.

We investigated the influence of the different number of measurements for the CBT on our method's uncertainty. We conclude that determining the correct coefficients becomes more challenging in the high dissociation region with large static correlation. A sufficient number of measurements must be performed to determine the CCSD amplitudes with sufficient accuracy. For the investigated molecules, a total of $R=100$, $N_{\text{sample}} =10^6$, $N_{U}=10^6$, and $N_{V}=10^6$ measurements were appropriate to predict energies with a high level of confidence. 

In addition, we applied our method with the perturbative triples correction to Cope-rearrangement, a well-known organic reaction. We demonstrated that our method produced activation energy comparable to MR-AQCC. This showed our method's potential for complex chemical reactions.

Nevertheless, for more complex systems or in the presence of real device errors such as depolarising noise, further verification is required to determine the extent to which QC-CBT-TCC works well. Although, in this study, we utilized the CBT method to approximate the wavefunction on the quantum computer, it is possible to explore extensions that combine TCC with methods that are more resilient to noise and statistical errors, such as quantum selected configuration interaction~\cite{kanno2023quantum,Nakagawa2023arXiv}. 
Another potential avenue for further development of this method is to create a self-consistent version~\cite{liao2024unveiling,feldmann2024complete} that iterates between the calculation of the active space wavefunction and the optimization of the coupled cluster amplitudes in the outside of the active space, although such a self-consistent approach multiplies the computational cost associated with the method.

Finally, the performance of a tomography method depends heavily on the number of shots available. Matchgate/Fermionic shadows has good asymptotic scaling and can be highly efficient if enough shots are available~\cite{zhao2021fermionic,wan2023matchgate}. On the other hand, these shadow tomography methods may not always be the best approach given realistic shot budgets~\cite{takemori2023balancing}.
We chose CBT for this study because it requires shallower circuits and can be effective with small shot budgets by adjusting the $R$ parameter. Nevertheless, it is an open question as to which method is better for extracting CI coefficients with a limited number of shots.

\begin{acknowledgments}
L.E was supported by the Japanese Government (MEXT) scholarship for his Ph.D. studies at Osaka University, Japan. 
L.E. also received support through a collaborative research project with AGC Inc. 
This project was supported by funding from the MEXT Quantum Leap Flagship Program (MEXTQLEAP) through Grant No. JPMXS0120319794, and the JST COI-NEXT Program through Grant No. JPMJPF2014. The completion of this research was partially facilitated by the JSPS Grants-in-Aid for Scientific Research (KAKENHI), specifically Grant Nos. JP23H03819 and JP21K18933.
\end{acknowledgments}

\appendix

\section{Number of shot dependency of QC-CBT-T for LiH}
\begin{figure*}
	\includegraphics[bb=0 0 989 327, width=\textwidth]{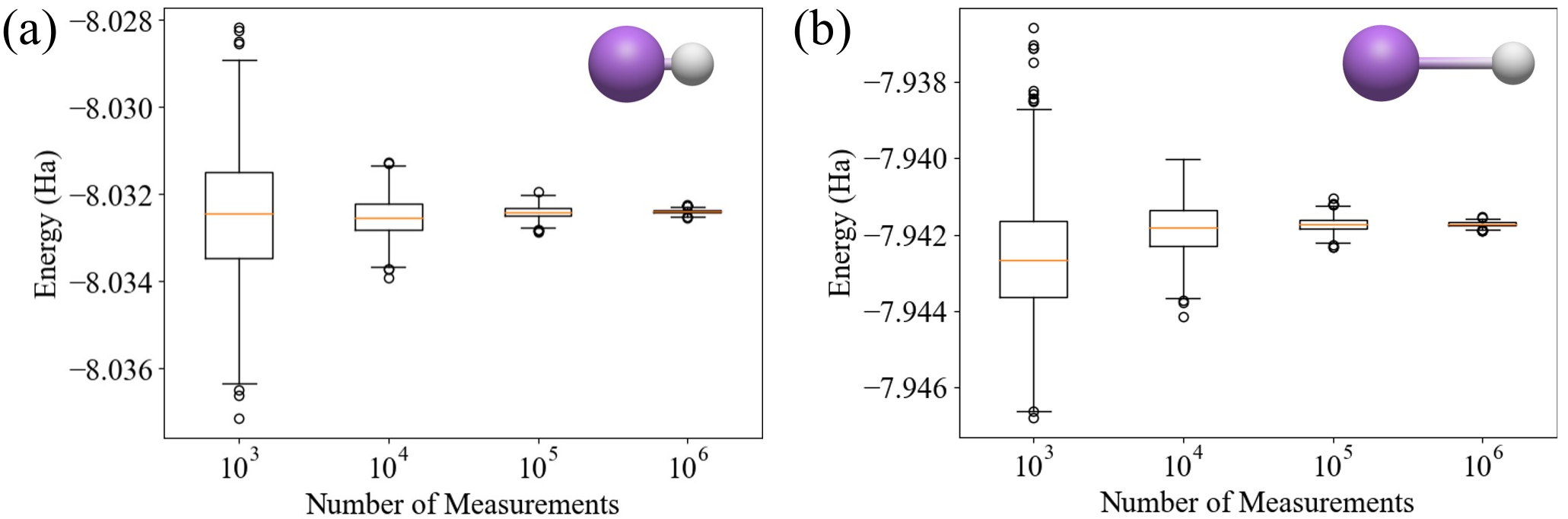}
	\caption{Energy insecurity of QC-CBT-TCC(c) for LiH with a different number of measurements for CBT. $N_{\text{sample}}, N_{U}$, and $N_{V}$ were all measured in the specified quantities in the figure as the X-axis values. CBT was performed 1000 times for each setting to obtain an estimate for the statistical errors. The box indicates the range from the first to the third quartile called the inter-quartile range (IQR), with the median drawn in the middle. The box contains $50 \% $ of the data. Whiskers are the lines extending 1.5 times the IQR from the first and third quantiles. Data points that exceed the whiskers are considered outliers (fliers) and are represented as single dots.  (a) Number of measurements dependency at an equilibrium distance of 3.016 Bohr. (b) Number of measurements dependency at 6.032 Bohr, twice the equilibrium distance.}
	\label{Figure_LiH_energy_uncertainty_equilibrium}
\end{figure*}

In Figures \ref{Figure_LiH_energy_uncertainty_equilibrium} (a) and \ref{Figure_LiH_energy_uncertainty_equilibrium} (b) the results for LiH show a behavior similar to that of H\textsubscript{2}O and N\textsubscript{2} in that increasing the number of measurements reduces the corrected energy distribution. This behavior can be explained by the higher accuracy and decreased uncertainty in determining the coefficients of a state on a quantum computer. Consequently, the obtained CBT state is more consistent, and therefore, the expected energy can be determined with a higher degree of confidence.

\bibliography{main.bib}

\begin{thebibliography}{91}%
\makeatletter
\providecommand \@ifxundefined [1]{%
 \@ifx{#1\undefined}
}%
\providecommand \@ifnum [1]{%
 \ifnum #1\expandafter \@firstoftwo
 \else \expandafter \@secondoftwo
 \fi
}%
\providecommand \@ifx [1]{%
 \ifx #1\expandafter \@firstoftwo
 \else \expandafter \@secondoftwo
 \fi
}%
\providecommand \natexlab [1]{#1}%
\providecommand \enquote  [1]{``#1''}%
\providecommand \bibnamefont  [1]{#1}%
\providecommand \bibfnamefont [1]{#1}%
\providecommand \citenamefont [1]{#1}%
\providecommand \href@noop [0]{\@secondoftwo}%
\providecommand \href [0]{\begingroup \@sanitize@url \@href}%
\providecommand \@href[1]{\@@startlink{#1}\@@href}%
\providecommand \@@href[1]{\endgroup#1\@@endlink}%
\providecommand \@sanitize@url [0]{\catcode `\\12\catcode `\$12\catcode
  `\&12\catcode `\#12\catcode `\^12\catcode `\_12\catcode `\%12\relax}%
\providecommand \@@startlink[1]{}%
\providecommand \@@endlink[0]{}%
\providecommand \url  [0]{\begingroup\@sanitize@url \@url }%
\providecommand \@url [1]{\endgroup\@href {#1}{\urlprefix }}%
\providecommand \urlprefix  [0]{URL }%
\providecommand \Eprint [0]{\href }%
\providecommand \doibase [0]{https://doi.org/}%
\providecommand \selectlanguage [0]{\@gobble}%
\providecommand \bibinfo  [0]{\@secondoftwo}%
\providecommand \bibfield  [0]{\@secondoftwo}%
\providecommand \translation [1]{[#1]}%
\providecommand \BibitemOpen [0]{}%
\providecommand \bibitemStop [0]{}%
\providecommand \bibitemNoStop [0]{.\EOS\space}%
\providecommand \EOS [0]{\spacefactor3000\relax}%
\providecommand \BibitemShut  [1]{\csname bibitem#1\endcsname}%
\let\auto@bib@innerbib\@empty
\bibitem [{\citenamefont {Cao}\ \emph {et~al.}(2019)\citenamefont {Cao},
  \citenamefont {Romero}, \citenamefont {Olson}, \citenamefont {Degroote},
  \citenamefont {Johnson}, \citenamefont {Kieferov{\'{a}}}, \citenamefont
  {Kivlichan}, \citenamefont {Menke}, \citenamefont {Peropadre}, \citenamefont
  {Sawaya}, \citenamefont {Sim}, \citenamefont {Veis},\ and\ \citenamefont
  {Aspuru-Guzik}}]{Cao2019}%
  \BibitemOpen
  \bibfield  {author} {\bibinfo {author} {\bibfnamefont {Y.}~\bibnamefont
  {Cao}}, \bibinfo {author} {\bibfnamefont {J.}~\bibnamefont {Romero}},
  \bibinfo {author} {\bibfnamefont {J.~P.}\ \bibnamefont {Olson}}, \bibinfo
  {author} {\bibfnamefont {M.}~\bibnamefont {Degroote}}, \bibinfo {author}
  {\bibfnamefont {P.~D.}\ \bibnamefont {Johnson}}, \bibinfo {author}
  {\bibfnamefont {M.}~\bibnamefont {Kieferov{\'{a}}}}, \bibinfo {author}
  {\bibfnamefont {I.~D.}\ \bibnamefont {Kivlichan}}, \bibinfo {author}
  {\bibfnamefont {T.}~\bibnamefont {Menke}}, \bibinfo {author} {\bibfnamefont
  {B.}~\bibnamefont {Peropadre}}, \bibinfo {author} {\bibfnamefont {N.~P.}\
  \bibnamefont {Sawaya}}, \bibinfo {author} {\bibfnamefont {S.}~\bibnamefont
  {Sim}}, \bibinfo {author} {\bibfnamefont {L.}~\bibnamefont {Veis}},\ and\
  \bibinfo {author} {\bibfnamefont {A.}~\bibnamefont {Aspuru-Guzik}},\ }\href
  {https://doi.org/10.1021/ACS.CHEMREV.8B00803} {\bibfield  {journal} {\bibinfo
   {journal} {Chemical Reviews}\ }\textbf {\bibinfo {volume} {119}},\ \bibinfo
  {pages} {10856} (\bibinfo {year} {2019})}\BibitemShut {NoStop}%
\bibitem [{\citenamefont {Babbush}\ \emph {et~al.}(2021)\citenamefont
  {Babbush}, \citenamefont {McClean}, \citenamefont {Newman}, \citenamefont
  {Gidney}, \citenamefont {Boixo},\ and\ \citenamefont
  {Neven}}]{Babbush2021PRXQ}%
  \BibitemOpen
  \bibfield  {author} {\bibinfo {author} {\bibfnamefont {R.}~\bibnamefont
  {Babbush}}, \bibinfo {author} {\bibfnamefont {J.~R.}\ \bibnamefont
  {McClean}}, \bibinfo {author} {\bibfnamefont {M.}~\bibnamefont {Newman}},
  \bibinfo {author} {\bibfnamefont {C.}~\bibnamefont {Gidney}}, \bibinfo
  {author} {\bibfnamefont {S.}~\bibnamefont {Boixo}},\ and\ \bibinfo {author}
  {\bibfnamefont {H.}~\bibnamefont {Neven}},\ }\href
  {https://journals.aps.org/prxquantum/abstract/10.1103/PRXQuantum.2.010103}
  {\bibfield  {journal} {\bibinfo  {journal} {PRX Quantum}\ }\textbf {\bibinfo
  {volume} {2}},\ \bibinfo {pages} {010103} (\bibinfo {year}
  {2021})}\BibitemShut {NoStop}%
\bibitem [{\citenamefont {Goings}\ \emph {et~al.}(2022)\citenamefont {Goings},
  \citenamefont {White}, \citenamefont {Lee}, \citenamefont {Tautermann},
  \citenamefont {Degroote}, \citenamefont {Gidney}, \citenamefont {Shiozaki},
  \citenamefont {Babbush},\ and\ \citenamefont {Rubin}}]{Goings2022PNAS}%
  \BibitemOpen
  \bibfield  {author} {\bibinfo {author} {\bibfnamefont {J.~J.}\ \bibnamefont
  {Goings}}, \bibinfo {author} {\bibfnamefont {A.}~\bibnamefont {White}},
  \bibinfo {author} {\bibfnamefont {J.}~\bibnamefont {Lee}}, \bibinfo {author}
  {\bibfnamefont {C.~S.}\ \bibnamefont {Tautermann}}, \bibinfo {author}
  {\bibfnamefont {M.}~\bibnamefont {Degroote}}, \bibinfo {author}
  {\bibfnamefont {C.}~\bibnamefont {Gidney}}, \bibinfo {author} {\bibfnamefont
  {T.}~\bibnamefont {Shiozaki}}, \bibinfo {author} {\bibfnamefont
  {R.}~\bibnamefont {Babbush}},\ and\ \bibinfo {author} {\bibfnamefont {N.~C.}\
  \bibnamefont {Rubin}},\ }\href
  {https://www.pnas.org/doi/10.1073/pnas.2203533119} {\bibfield  {journal}
  {\bibinfo  {journal} {Proceedings of the National Academy of Sciences}\
  }\textbf {\bibinfo {volume} {119}},\ \bibinfo {pages} {e2203533119} (\bibinfo
  {year} {2022})}\BibitemShut {NoStop}%
\bibitem [{\citenamefont {Stein}\ and\ \citenamefont
  {Reiher}(2016)}]{Stein2016}%
  \BibitemOpen
  \bibfield  {author} {\bibinfo {author} {\bibfnamefont {C.~J.}\ \bibnamefont
  {Stein}}\ and\ \bibinfo {author} {\bibfnamefont {M.}~\bibnamefont {Reiher}},\
  }\href
  {https://doi.org/10.1021/ACS.JCTC.6B00156/ASSET/IMAGES/LARGE/CT-2016-00156G_0011.JPEG}
  {\bibfield  {journal} {\bibinfo  {journal} {Journal of Chemical Theory and
  Computation}\ }\textbf {\bibinfo {volume} {12}},\ \bibinfo {pages} {1760}
  (\bibinfo {year} {2016})},\ \Eprint {https://arxiv.org/abs/1602.03835}
  {arXiv:1602.03835} \BibitemShut {NoStop}%
\bibitem [{\citenamefont {Sayfutyarova}\ \emph {et~al.}(2017)\citenamefont
  {Sayfutyarova}, \citenamefont {Sun}, \citenamefont {Chan},\ and\
  \citenamefont {Knizia}}]{sayfutyarova2017automated}%
  \BibitemOpen
  \bibfield  {author} {\bibinfo {author} {\bibfnamefont {E.~R.}\ \bibnamefont
  {Sayfutyarova}}, \bibinfo {author} {\bibfnamefont {Q.}~\bibnamefont {Sun}},
  \bibinfo {author} {\bibfnamefont {G.~K.-L.}\ \bibnamefont {Chan}},\ and\
  \bibinfo {author} {\bibfnamefont {G.}~\bibnamefont {Knizia}},\ }\href@noop {}
  {\bibfield  {journal} {\bibinfo  {journal} {Journal of chemical theory and
  computation}\ }\textbf {\bibinfo {volume} {13}},\ \bibinfo {pages} {4063}
  (\bibinfo {year} {2017})}\BibitemShut {NoStop}%
\bibitem [{\citenamefont {Bao}\ \emph {et~al.}(2018)\citenamefont {Bao},
  \citenamefont {Dong}, \citenamefont {Gagliardi},\ and\ \citenamefont
  {Truhlar}}]{bao2018automatic}%
  \BibitemOpen
  \bibfield  {author} {\bibinfo {author} {\bibfnamefont {J.~J.}\ \bibnamefont
  {Bao}}, \bibinfo {author} {\bibfnamefont {S.~S.}\ \bibnamefont {Dong}},
  \bibinfo {author} {\bibfnamefont {L.}~\bibnamefont {Gagliardi}},\ and\
  \bibinfo {author} {\bibfnamefont {D.~G.}\ \bibnamefont {Truhlar}},\
  }\href@noop {} {\bibfield  {journal} {\bibinfo  {journal} {Journal of
  Chemical Theory and Computation}\ }\textbf {\bibinfo {volume} {14}},\
  \bibinfo {pages} {2017} (\bibinfo {year} {2018})}\BibitemShut {NoStop}%
\bibitem [{\citenamefont {Sayfutyarova}\ and\ \citenamefont
  {Hammes-Schiffer}(2019)}]{sayfutyarova2019constructing}%
  \BibitemOpen
  \bibfield  {author} {\bibinfo {author} {\bibfnamefont {E.~R.}\ \bibnamefont
  {Sayfutyarova}}\ and\ \bibinfo {author} {\bibfnamefont {S.}~\bibnamefont
  {Hammes-Schiffer}},\ }\href@noop {} {\bibfield  {journal} {\bibinfo
  {journal} {Journal of chemical theory and computation}\ }\textbf {\bibinfo
  {volume} {15}},\ \bibinfo {pages} {1679} (\bibinfo {year}
  {2019})}\BibitemShut {NoStop}%
\bibitem [{\citenamefont {Jeong}\ \emph {et~al.}(2020)\citenamefont {Jeong},
  \citenamefont {Stoneburner}, \citenamefont {King}, \citenamefont {Li},
  \citenamefont {Walker}, \citenamefont {Lindh},\ and\ \citenamefont
  {Gagliardi}}]{jeong2020automation}%
  \BibitemOpen
  \bibfield  {author} {\bibinfo {author} {\bibfnamefont {W.}~\bibnamefont
  {Jeong}}, \bibinfo {author} {\bibfnamefont {S.~J.}\ \bibnamefont
  {Stoneburner}}, \bibinfo {author} {\bibfnamefont {D.}~\bibnamefont {King}},
  \bibinfo {author} {\bibfnamefont {R.}~\bibnamefont {Li}}, \bibinfo {author}
  {\bibfnamefont {A.}~\bibnamefont {Walker}}, \bibinfo {author} {\bibfnamefont
  {R.}~\bibnamefont {Lindh}},\ and\ \bibinfo {author} {\bibfnamefont
  {L.}~\bibnamefont {Gagliardi}},\ }\href@noop {} {\bibfield  {journal}
  {\bibinfo  {journal} {Journal of Chemical Theory and Computation}\ }\textbf
  {\bibinfo {volume} {16}},\ \bibinfo {pages} {2389} (\bibinfo {year}
  {2020})}\BibitemShut {NoStop}%
\bibitem [{\citenamefont {King}\ and\ \citenamefont
  {Gagliardi}(2021)}]{king2021ranked}%
  \BibitemOpen
  \bibfield  {author} {\bibinfo {author} {\bibfnamefont {D.~S.}\ \bibnamefont
  {King}}\ and\ \bibinfo {author} {\bibfnamefont {L.}~\bibnamefont
  {Gagliardi}},\ }\href@noop {} {\bibfield  {journal} {\bibinfo  {journal}
  {Journal of Chemical Theory and Computation}\ }\textbf {\bibinfo {volume}
  {17}},\ \bibinfo {pages} {2817} (\bibinfo {year} {2021})}\BibitemShut
  {NoStop}%
\bibitem [{\citenamefont {King}\ \emph {et~al.}(2022)\citenamefont {King},
  \citenamefont {Hermes}, \citenamefont {Truhlar},\ and\ \citenamefont
  {Gagliardi}}]{king2022large}%
  \BibitemOpen
  \bibfield  {author} {\bibinfo {author} {\bibfnamefont {D.~S.}\ \bibnamefont
  {King}}, \bibinfo {author} {\bibfnamefont {M.~R.}\ \bibnamefont {Hermes}},
  \bibinfo {author} {\bibfnamefont {D.~G.}\ \bibnamefont {Truhlar}},\ and\
  \bibinfo {author} {\bibfnamefont {L.}~\bibnamefont {Gagliardi}},\ }\href@noop
  {} {\bibfield  {journal} {\bibinfo  {journal} {Journal of Chemical Theory and
  Computation}\ }\textbf {\bibinfo {volume} {18}},\ \bibinfo {pages} {6065}
  (\bibinfo {year} {2022})}\BibitemShut {NoStop}%
\bibitem [{\citenamefont {Kaufold}\ \emph {et~al.}(2023)\citenamefont
  {Kaufold}, \citenamefont {Chintala}, \citenamefont {Pandeya},\ and\
  \citenamefont {Dong}}]{kaufold2023automated}%
  \BibitemOpen
  \bibfield  {author} {\bibinfo {author} {\bibfnamefont {B.~W.}\ \bibnamefont
  {Kaufold}}, \bibinfo {author} {\bibfnamefont {N.}~\bibnamefont {Chintala}},
  \bibinfo {author} {\bibfnamefont {P.}~\bibnamefont {Pandeya}},\ and\ \bibinfo
  {author} {\bibfnamefont {S.~S.}\ \bibnamefont {Dong}},\ }\href@noop {}
  {\bibfield  {journal} {\bibinfo  {journal} {Journal of Chemical Theory and
  Computation}\ }\textbf {\bibinfo {volume} {19}},\ \bibinfo {pages} {2469}
  (\bibinfo {year} {2023})}\BibitemShut {NoStop}%
\bibitem [{\citenamefont {Lyakh}\ \emph {et~al.}(2012)\citenamefont {Lyakh},
  \citenamefont {Musia{\l}}, \citenamefont {Lotrich},\ and\ \citenamefont
  {Bartlett}}]{Lyakh2012}%
  \BibitemOpen
  \bibfield  {author} {\bibinfo {author} {\bibfnamefont {D.~I.}\ \bibnamefont
  {Lyakh}}, \bibinfo {author} {\bibfnamefont {M.}~\bibnamefont {Musia{\l}}},
  \bibinfo {author} {\bibfnamefont {V.~F.}\ \bibnamefont {Lotrich}},\ and\
  \bibinfo {author} {\bibfnamefont {R.~J.}\ \bibnamefont {Bartlett}},\ }\href
  {https://pubs.acs.org/doi/full/10.1021/cr2001417} {\bibfield  {journal}
  {\bibinfo  {journal} {Chemical Reviews}\ }\textbf {\bibinfo {volume} {112}},\
  \bibinfo {pages} {182} (\bibinfo {year} {2012})}\BibitemShut {NoStop}%
\bibitem [{\citenamefont {Yanai}\ \emph {et~al.}(2015)\citenamefont {Yanai},
  \citenamefont {Kurashige}, \citenamefont {Mizukami}, \citenamefont
  {Chalupsk{\'{y}}}, \citenamefont {Lan},\ and\ \citenamefont
  {Saitow}}]{Yanai2015}%
  \BibitemOpen
  \bibfield  {author} {\bibinfo {author} {\bibfnamefont {T.}~\bibnamefont
  {Yanai}}, \bibinfo {author} {\bibfnamefont {Y.}~\bibnamefont {Kurashige}},
  \bibinfo {author} {\bibfnamefont {W.}~\bibnamefont {Mizukami}}, \bibinfo
  {author} {\bibfnamefont {J.}~\bibnamefont {Chalupsk{\'{y}}}}, \bibinfo
  {author} {\bibfnamefont {T.~N.}\ \bibnamefont {Lan}},\ and\ \bibinfo {author}
  {\bibfnamefont {M.}~\bibnamefont {Saitow}},\ }\href
  {https://doi.org/10.1002/QUA.24808} {\bibfield  {journal} {\bibinfo
  {journal} {International Journal of Quantum Chemistry}\ }\textbf {\bibinfo
  {volume} {115}},\ \bibinfo {pages} {283} (\bibinfo {year}
  {2015})}\BibitemShut {NoStop}%
\bibitem [{\citenamefont {Roos}\ \emph {et~al.}(2016)\citenamefont {Roos},
  \citenamefont {Lindh}, \citenamefont {Malmqvist}, \citenamefont {Veryazov},\
  and\ \citenamefont {Widmark}}]{Roos2016}%
  \BibitemOpen
  \bibfield  {author} {\bibinfo {author} {\bibfnamefont {B.~O.}\ \bibnamefont
  {Roos}}, \bibinfo {author} {\bibfnamefont {R.}~\bibnamefont {Lindh}},
  \bibinfo {author} {\bibfnamefont {P.}~\bibnamefont {Malmqvist}}, \bibinfo
  {author} {\bibfnamefont {V.}~\bibnamefont {Veryazov}},\ and\ \bibinfo
  {author} {\bibfnamefont {P.~O.}\ \bibnamefont {Widmark}},\ }\href
  {https://doi.org/10.1002/9781119126171} {\bibfield  {journal} {\bibinfo
  {journal} {Multiconfigurational Quantum Chemistry}\ ,\ \bibinfo {pages} {1}}
  (\bibinfo {year} {2016})}\BibitemShut {NoStop}%
\bibitem [{\citenamefont {Pathak}\ \emph {et~al.}(2017)\citenamefont {Pathak},
  \citenamefont {Lang},\ and\ \citenamefont {Neese}}]{Pathak2017}%
  \BibitemOpen
  \bibfield  {author} {\bibinfo {author} {\bibfnamefont {S.}~\bibnamefont
  {Pathak}}, \bibinfo {author} {\bibfnamefont {L.}~\bibnamefont {Lang}},\ and\
  \bibinfo {author} {\bibfnamefont {F.}~\bibnamefont {Neese}},\ }\href
  {https://doi.org/10.1063/1.5017942} {\bibfield  {journal} {\bibinfo
  {journal} {Journal of Chemical Physics}\ }\textbf {\bibinfo {volume} {147}},\
  \bibinfo {pages} {234109} (\bibinfo {year} {2017})}\BibitemShut {NoStop}%
\bibitem [{\citenamefont {Park}\ \emph {et~al.}(2020)\citenamefont {Park},
  \citenamefont {Al-Saadon}, \citenamefont {Macleod}, \citenamefont
  {Shiozaki},\ and\ \citenamefont {Vlaisavljevich}}]{Park2020}%
  \BibitemOpen
  \bibfield  {author} {\bibinfo {author} {\bibfnamefont {J.~W.}\ \bibnamefont
  {Park}}, \bibinfo {author} {\bibfnamefont {R.}~\bibnamefont {Al-Saadon}},
  \bibinfo {author} {\bibfnamefont {M.~K.}\ \bibnamefont {Macleod}}, \bibinfo
  {author} {\bibfnamefont {T.}~\bibnamefont {Shiozaki}},\ and\ \bibinfo
  {author} {\bibfnamefont {B.}~\bibnamefont {Vlaisavljevich}},\ }\href
  {https://pubs.acs.org/doi/full/10.1021/acs.chemrev.9b00496} {\bibfield
  {journal} {\bibinfo  {journal} {Chemical Reviews}\ }\textbf {\bibinfo
  {volume} {120}},\ \bibinfo {pages} {5878} (\bibinfo {year} {2020})},\ \Eprint
  {https://arxiv.org/abs/1911.06836} {arXiv:1911.06836} \BibitemShut {NoStop}%
\bibitem [{\citenamefont {Khedkar}\ and\ \citenamefont
  {Roemelt}(2021)}]{Khedkar2021}%
  \BibitemOpen
  \bibfield  {author} {\bibinfo {author} {\bibfnamefont {A.}~\bibnamefont
  {Khedkar}}\ and\ \bibinfo {author} {\bibfnamefont {M.}~\bibnamefont
  {Roemelt}},\ }\href {https://doi.org/10.1039/D1CP02640B} {\bibfield
  {journal} {\bibinfo  {journal} {Physical Chemistry Chemical Physics}\
  }\textbf {\bibinfo {volume} {23}},\ \bibinfo {pages} {17097} (\bibinfo {year}
  {2021})}\BibitemShut {NoStop}%
\bibitem [{\citenamefont {Casanova}(2022)}]{Casanova2022}%
  \BibitemOpen
  \bibfield  {author} {\bibinfo {author} {\bibfnamefont {D.}~\bibnamefont
  {Casanova}},\ }\href {https://doi.org/10.1002/WCMS.1561} {\bibfield
  {journal} {\bibinfo  {journal} {Wiley Interdisciplinary Reviews:
  Computational Molecular Science}\ }\textbf {\bibinfo {volume} {12}},\
  \bibinfo {pages} {e1561} (\bibinfo {year} {2022})}\BibitemShut {NoStop}%
\bibitem [{\citenamefont {McArdle}\ and\ \citenamefont
  {Tew}(2020)}]{McArdle2020}%
  \BibitemOpen
  \bibfield  {author} {\bibinfo {author} {\bibfnamefont {S.}~\bibnamefont
  {McArdle}}\ and\ \bibinfo {author} {\bibfnamefont {D.~P.}\ \bibnamefont
  {Tew}},\ }\href {https://arxiv.org/abs/2006.11181v1} {\bibfield  {journal}
  {\bibinfo  {journal} {arXiv}\ } (\bibinfo {year} {2020})},\ \Eprint
  {https://arxiv.org/abs/2006.11181} {arXiv:2006.11181} \BibitemShut {NoStop}%
\bibitem [{\citenamefont {Kumar}\ \emph {et~al.}(2022)\citenamefont {Kumar},
  \citenamefont {Asthana}, \citenamefont {Masteran}, \citenamefont {Valeev},
  \citenamefont {Zhang}, \citenamefont {Cincio}, \citenamefont {Tretiak},\ and\
  \citenamefont {Dub}}]{Kumar2022}%
  \BibitemOpen
  \bibfield  {author} {\bibinfo {author} {\bibfnamefont {A.}~\bibnamefont
  {Kumar}}, \bibinfo {author} {\bibfnamefont {A.}~\bibnamefont {Asthana}},
  \bibinfo {author} {\bibfnamefont {C.}~\bibnamefont {Masteran}}, \bibinfo
  {author} {\bibfnamefont {E.~F.}\ \bibnamefont {Valeev}}, \bibinfo {author}
  {\bibfnamefont {Y.}~\bibnamefont {Zhang}}, \bibinfo {author} {\bibfnamefont
  {L.}~\bibnamefont {Cincio}}, \bibinfo {author} {\bibfnamefont
  {S.}~\bibnamefont {Tretiak}},\ and\ \bibinfo {author} {\bibfnamefont {P.~A.}\
  \bibnamefont {Dub}},\ }\href
  {https://pubs.acs.org/doi/full/10.1021/acs.jctc.2c00520} {\bibfield
  {journal} {\bibinfo  {journal} {Journal of Chemical Theory and Computation}\
  }\textbf {\bibinfo {volume} {18}},\ \bibinfo {pages} {5312} (\bibinfo {year}
  {2022})}\BibitemShut {NoStop}%
\bibitem [{\citenamefont {Sokolov}\ \emph {et~al.}(2023)\citenamefont
  {Sokolov}, \citenamefont {Dobrautz}, \citenamefont {Luo}, \citenamefont
  {Alavi},\ and\ \citenamefont {Tavernelli}}]{Sokolov2023}%
  \BibitemOpen
  \bibfield  {author} {\bibinfo {author} {\bibfnamefont {I.~O.}\ \bibnamefont
  {Sokolov}}, \bibinfo {author} {\bibfnamefont {W.}~\bibnamefont {Dobrautz}},
  \bibinfo {author} {\bibfnamefont {H.}~\bibnamefont {Luo}}, \bibinfo {author}
  {\bibfnamefont {A.}~\bibnamefont {Alavi}},\ and\ \bibinfo {author}
  {\bibfnamefont {I.}~\bibnamefont {Tavernelli}},\ }\href
  {https://doi.org/10.1103/PhysRevResearch.5.023174} {\bibfield  {journal}
  {\bibinfo  {journal} {Phys. Rev. Res.}\ }\textbf {\bibinfo {volume} {5}},\
  \bibinfo {pages} {023174} (\bibinfo {year} {2023})}\BibitemShut {NoStop}%
\bibitem [{\citenamefont {Bauman}\ \emph {et~al.}(2019)\citenamefont {Bauman},
  \citenamefont {Bylaska}, \citenamefont {Krishnamoorthy}, \citenamefont {Low},
  \citenamefont {Wiebe}, \citenamefont {Granade}, \citenamefont {Roetteler},
  \citenamefont {Troyer},\ and\ \citenamefont {Kowalski}}]{Bauman2019}%
  \BibitemOpen
  \bibfield  {author} {\bibinfo {author} {\bibfnamefont {N.~P.}\ \bibnamefont
  {Bauman}}, \bibinfo {author} {\bibfnamefont {E.~J.}\ \bibnamefont {Bylaska}},
  \bibinfo {author} {\bibfnamefont {S.}~\bibnamefont {Krishnamoorthy}},
  \bibinfo {author} {\bibfnamefont {G.~H.}\ \bibnamefont {Low}}, \bibinfo
  {author} {\bibfnamefont {N.}~\bibnamefont {Wiebe}}, \bibinfo {author}
  {\bibfnamefont {C.~E.}\ \bibnamefont {Granade}}, \bibinfo {author}
  {\bibfnamefont {M.}~\bibnamefont {Roetteler}}, \bibinfo {author}
  {\bibfnamefont {M.}~\bibnamefont {Troyer}},\ and\ \bibinfo {author}
  {\bibfnamefont {K.}~\bibnamefont {Kowalski}},\ }\href
  {https://doi.org/10.1063/1.5094643} {\bibfield  {journal} {\bibinfo
  {journal} {Journal of Chemical Physics}\ }\textbf {\bibinfo {volume} {151}},\
  \bibinfo {pages} {14107} (\bibinfo {year} {2019})},\ \Eprint
  {https://arxiv.org/abs/1902.01553} {arXiv:1902.01553} \BibitemShut {NoStop}%
\bibitem [{\citenamefont {Metcalf}\ \emph {et~al.}(2020)\citenamefont
  {Metcalf}, \citenamefont {Bauman}, \citenamefont {Kowalski},\ and\
  \citenamefont {{De Jong}}}]{Metcalf2020}%
  \BibitemOpen
  \bibfield  {author} {\bibinfo {author} {\bibfnamefont {M.}~\bibnamefont
  {Metcalf}}, \bibinfo {author} {\bibfnamefont {N.~P.}\ \bibnamefont {Bauman}},
  \bibinfo {author} {\bibfnamefont {K.}~\bibnamefont {Kowalski}},\ and\
  \bibinfo {author} {\bibfnamefont {W.~A.}\ \bibnamefont {{De Jong}}},\ }\href
  {https://doi.org/10.1021/ACS.JCTC.0C00421} {\bibfield  {journal} {\bibinfo
  {journal} {Journal of Chemical Theory and Computation}\ }\textbf {\bibinfo
  {volume} {16}},\ \bibinfo {pages} {6165} (\bibinfo {year} {2020})},\ \Eprint
  {https://arxiv.org/abs/2004.07721} {arXiv:2004.07721} \BibitemShut {NoStop}%
\bibitem [{\citenamefont {{Nicholas P.}}\ \emph {et~al.}(2021)\citenamefont
  {{Nicholas P.}}, \citenamefont {Chl{\'{a}}dek}, \citenamefont {Veis},
  \citenamefont {Pittner},\ and\ \citenamefont {Karol}}]{NicholasP2021}%
  \BibitemOpen
  \bibfield  {author} {\bibinfo {author} {\bibfnamefont {B.}~\bibnamefont
  {{Nicholas P.}}}, \bibinfo {author} {\bibfnamefont {J.}~\bibnamefont
  {Chl{\'{a}}dek}}, \bibinfo {author} {\bibfnamefont {L.}~\bibnamefont {Veis}},
  \bibinfo {author} {\bibfnamefont {J.}~\bibnamefont {Pittner}},\ and\ \bibinfo
  {author} {\bibfnamefont {K.}~\bibnamefont {Karol}},\ }\href
  {https://doi.org/10.1088/2058-9565/ABF602} {\bibfield  {journal} {\bibinfo
  {journal} {Quantum Science and Technology}\ }\textbf {\bibinfo {volume}
  {6}},\ \bibinfo {pages} {034008} (\bibinfo {year} {2021})},\ \Eprint
  {https://arxiv.org/abs/2011.01985} {arXiv:2011.01985} \BibitemShut {NoStop}%
\bibitem [{\citenamefont {Huang}\ \emph {et~al.}(2023)\citenamefont {Huang},
  \citenamefont {Li},\ and\ \citenamefont {Evangelista}}]{Huang2022}%
  \BibitemOpen
  \bibfield  {author} {\bibinfo {author} {\bibfnamefont {R.}~\bibnamefont
  {Huang}}, \bibinfo {author} {\bibfnamefont {C.}~\bibnamefont {Li}},\ and\
  \bibinfo {author} {\bibfnamefont {F.~A.}\ \bibnamefont {Evangelista}},\
  }\href {https://doi.org/10.1103/PRXQuantum.4.020313} {\bibfield  {journal}
  {\bibinfo  {journal} {PRX Quantum}\ }\textbf {\bibinfo {volume} {4}},\
  \bibinfo {pages} {020313} (\bibinfo {year} {2023})},\ \Eprint
  {https://arxiv.org/abs/2208.08591} {arXiv:2208.08591} \BibitemShut {NoStop}%
\bibitem [{\citenamefont {Bauman}\ and\ \citenamefont
  {Kowalski}(2022)}]{Baumann2022}%
  \BibitemOpen
  \bibfield  {author} {\bibinfo {author} {\bibfnamefont {N.~P.}\ \bibnamefont
  {Bauman}}\ and\ \bibinfo {author} {\bibfnamefont {K.}~\bibnamefont
  {Kowalski}},\ }\href {https://doi.org/10.1186/S41313-022-00046-8} {\bibfield
  {journal} {\bibinfo  {journal} {Materials Theory 2022 6:1}\ }\textbf
  {\bibinfo {volume} {6}},\ \bibinfo {pages} {1} (\bibinfo {year}
  {2022})}\BibitemShut {NoStop}%
\bibitem [{\citenamefont {Le}\ and\ \citenamefont {Tran}(2023)}]{Le2023}%
  \BibitemOpen
  \bibfield  {author} {\bibinfo {author} {\bibfnamefont {N.~T.}\ \bibnamefont
  {Le}}\ and\ \bibinfo {author} {\bibfnamefont {L.~N.}\ \bibnamefont {Tran}},\
  }\href {https://doi.org/10.1021/ACS.JPCA.3C00993} {\bibfield  {journal}
  {\bibinfo  {journal} {The Journal of Physical Chemistry A}\ }\textbf
  {\bibinfo {volume} {127}},\ \bibinfo {pages} {5222} (\bibinfo {year}
  {2023})}\BibitemShut {NoStop}%
\bibitem [{\citenamefont {Ma}\ \emph {et~al.}(2021)\citenamefont {Ma},
  \citenamefont {Sheng}, \citenamefont {Govoni},\ and\ \citenamefont
  {Galli}}]{Ma2021JCTC}%
  \BibitemOpen
  \bibfield  {author} {\bibinfo {author} {\bibfnamefont {H.}~\bibnamefont
  {Ma}}, \bibinfo {author} {\bibfnamefont {N.}~\bibnamefont {Sheng}}, \bibinfo
  {author} {\bibfnamefont {M.}~\bibnamefont {Govoni}},\ and\ \bibinfo {author}
  {\bibfnamefont {G.}~\bibnamefont {Galli}},\ }\href
  {https://pubs.acs.org/doi/10.1021/acs.jctc.0c01258} {\bibfield  {journal}
  {\bibinfo  {journal} {Journal of Chemical Theory and Computation}\ }\textbf
  {\bibinfo {volume} {17}},\ \bibinfo {pages} {2116} (\bibinfo {year}
  {2021})}\BibitemShut {NoStop}%
\bibitem [{\citenamefont {Tammaro}\ \emph {et~al.}(2023)\citenamefont
  {Tammaro}, \citenamefont {Galli}, \citenamefont {Rice},\ and\ \citenamefont
  {Motta}}]{Tammaro2023}%
  \BibitemOpen
  \bibfield  {author} {\bibinfo {author} {\bibfnamefont {A.}~\bibnamefont
  {Tammaro}}, \bibinfo {author} {\bibfnamefont {D.~E.}\ \bibnamefont {Galli}},
  \bibinfo {author} {\bibfnamefont {J.~E.}\ \bibnamefont {Rice}},\ and\
  \bibinfo {author} {\bibfnamefont {M.}~\bibnamefont {Motta}},\ }\href
  {https://pubs.acs.org/doi/full/10.1021/acs.jpca.2c07653} {\bibfield
  {journal} {\bibinfo  {journal} {Journal of Physical Chemistry A}\ }\textbf
  {\bibinfo {volume} {127}},\ \bibinfo {pages} {817} (\bibinfo {year}
  {2023})},\ \Eprint {https://arxiv.org/abs/2202.13002} {arXiv:2202.13002}
  \BibitemShut {NoStop}%
\bibitem [{\citenamefont {Nishio}\ \emph {et~al.}(2023)\citenamefont {Nishio},
  \citenamefont {Oba},\ and\ \citenamefont {Kurashige}}]{Nishio2023PCCP}%
  \BibitemOpen
  \bibfield  {author} {\bibinfo {author} {\bibfnamefont {S.}~\bibnamefont
  {Nishio}}, \bibinfo {author} {\bibfnamefont {Y.}~\bibnamefont {Oba}},\ and\
  \bibinfo {author} {\bibfnamefont {Y.}~\bibnamefont {Kurashige}},\ }\href
  {https://pubs.rsc.org/en/content/articlelanding/2023/cp/d3cp03520d}
  {\bibfield  {journal} {\bibinfo  {journal} {Physical Chemistry Chemical
  Physics}\ }\textbf {\bibinfo {volume} {25}},\ \bibinfo {pages} {30525}
  (\bibinfo {year} {2023})}\BibitemShut {NoStop}%
\bibitem [{\citenamefont {Takeshita}\ \emph {et~al.}(2020)\citenamefont
  {Takeshita}, \citenamefont {Rubin}, \citenamefont {Jiang}, \citenamefont
  {Lee}, \citenamefont {Babbush},\ and\ \citenamefont
  {McClean}}]{Takeshita2020}%
  \BibitemOpen
  \bibfield  {author} {\bibinfo {author} {\bibfnamefont {T.}~\bibnamefont
  {Takeshita}}, \bibinfo {author} {\bibfnamefont {N.~C.}\ \bibnamefont
  {Rubin}}, \bibinfo {author} {\bibfnamefont {Z.}~\bibnamefont {Jiang}},
  \bibinfo {author} {\bibfnamefont {E.}~\bibnamefont {Lee}}, \bibinfo {author}
  {\bibfnamefont {R.}~\bibnamefont {Babbush}},\ and\ \bibinfo {author}
  {\bibfnamefont {J.~R.}\ \bibnamefont {McClean}},\ }\href
  {https://journals.aps.org/prx/abstract/10.1103/PhysRevX.10.011004} {\bibfield
   {journal} {\bibinfo  {journal} {Physical Review X}\ }\textbf {\bibinfo
  {volume} {10}},\ \bibinfo {pages} {011004} (\bibinfo {year} {2020})},\
  \Eprint {https://arxiv.org/abs/1902.10679} {arXiv:1902.10679} \BibitemShut
  {NoStop}%
\bibitem [{\citenamefont {Mizukami}\ \emph {et~al.}(2020)\citenamefont
  {Mizukami}, \citenamefont {Mitarai}, \citenamefont {Nakagawa}, \citenamefont
  {Yamamoto}, \citenamefont {Yan},\ and\ \citenamefont
  {Ohnishi}}]{Mizukami2020PRR}%
  \BibitemOpen
  \bibfield  {author} {\bibinfo {author} {\bibfnamefont {W.}~\bibnamefont
  {Mizukami}}, \bibinfo {author} {\bibfnamefont {K.}~\bibnamefont {Mitarai}},
  \bibinfo {author} {\bibfnamefont {Y.~O.}\ \bibnamefont {Nakagawa}}, \bibinfo
  {author} {\bibfnamefont {T.}~\bibnamefont {Yamamoto}}, \bibinfo {author}
  {\bibfnamefont {T.}~\bibnamefont {Yan}},\ and\ \bibinfo {author}
  {\bibfnamefont {Y.-y.}\ \bibnamefont {Ohnishi}},\ }\href
  {https://journals.aps.org/prresearch/abstract/10.1103/PhysRevResearch.2.033421}
  {\bibfield  {journal} {\bibinfo  {journal} {Physical Review Research}\
  }\textbf {\bibinfo {volume} {2}},\ \bibinfo {pages} {033421} (\bibinfo {year}
  {2020})}\BibitemShut {NoStop}%
\bibitem [{\citenamefont {Sokolov}\ \emph {et~al.}(2019)\citenamefont
  {Sokolov}, \citenamefont {Barkoutsos}, \citenamefont {Ollitrault},
  \citenamefont {Greenberg}, \citenamefont {Rice}, \citenamefont {Pistoia},\
  and\ \citenamefont {Tavernelli}}]{Sokolov2019}%
  \BibitemOpen
  \bibfield  {author} {\bibinfo {author} {\bibfnamefont {I.~O.}\ \bibnamefont
  {Sokolov}}, \bibinfo {author} {\bibfnamefont {P.~K.}\ \bibnamefont
  {Barkoutsos}}, \bibinfo {author} {\bibfnamefont {P.~J.}\ \bibnamefont
  {Ollitrault}}, \bibinfo {author} {\bibfnamefont {D.}~\bibnamefont
  {Greenberg}}, \bibinfo {author} {\bibfnamefont {J.}~\bibnamefont {Rice}},
  \bibinfo {author} {\bibfnamefont {M.}~\bibnamefont {Pistoia}},\ and\ \bibinfo
  {author} {\bibfnamefont {I.}~\bibnamefont {Tavernelli}},\ }\bibfield
  {journal} {\bibinfo  {journal} {Journal of Chemical Physics}\ }\textbf
  {\bibinfo {volume} {152}},\ \href {https://doi.org/10.1063/1.5141835}
  {10.1063/1.5141835} (\bibinfo {year} {2019}),\ \Eprint
  {https://arxiv.org/abs/1911.10864v2} {arXiv:1911.10864v2} \BibitemShut
  {NoStop}%
\bibitem [{\citenamefont {Kottmann}\ \emph {et~al.}(2021)\citenamefont
  {Kottmann}, \citenamefont {Schleich}, \citenamefont {Tamayo-Mendoza},\ and\
  \citenamefont {Aspuru-Guzik}}]{Kottmann2021}%
  \BibitemOpen
  \bibfield  {author} {\bibinfo {author} {\bibfnamefont {J.~S.}\ \bibnamefont
  {Kottmann}}, \bibinfo {author} {\bibfnamefont {P.}~\bibnamefont {Schleich}},
  \bibinfo {author} {\bibfnamefont {T.}~\bibnamefont {Tamayo-Mendoza}},\ and\
  \bibinfo {author} {\bibfnamefont {A.}~\bibnamefont {Aspuru-Guzik}},\ }\href
  {https://pubs.acs.org/doi/full/10.1021/acs.jpclett.0c03410} {\bibfield
  {journal} {\bibinfo  {journal} {Journal of Physical Chemistry Letters}\
  }\textbf {\bibinfo {volume} {12}},\ \bibinfo {pages} {663} (\bibinfo {year}
  {2021})},\ \Eprint {https://arxiv.org/abs/2008.02819} {arXiv:2008.02819}
  \BibitemShut {NoStop}%
\bibitem [{\citenamefont {Schleich}\ \emph {et~al.}(2022)\citenamefont
  {Schleich}, \citenamefont {Kottmann},\ and\ \citenamefont
  {Aspuru-Guzik}}]{Schleich2022}%
  \BibitemOpen
  \bibfield  {author} {\bibinfo {author} {\bibfnamefont {P.}~\bibnamefont
  {Schleich}}, \bibinfo {author} {\bibfnamefont {J.~S.}\ \bibnamefont
  {Kottmann}},\ and\ \bibinfo {author} {\bibfnamefont {A.}~\bibnamefont
  {Aspuru-Guzik}},\ }\href {https://doi.org/10.1039/D2CP00247G} {\bibfield
  {journal} {\bibinfo  {journal} {Physical Chemistry Chemical Physics}\
  }\textbf {\bibinfo {volume} {24}},\ \bibinfo {pages} {13550} (\bibinfo {year}
  {2022})},\ \Eprint {https://arxiv.org/abs/2110.06812} {arXiv:2110.06812}
  \BibitemShut {NoStop}%
\bibitem [{\citenamefont {Rossmannek}\ \emph {et~al.}(2021)\citenamefont
  {Rossmannek}, \citenamefont {Barkoutsos}, \citenamefont {Ollitrault},\ and\
  \citenamefont {Tavernelli}}]{Rossmannek2021}%
  \BibitemOpen
  \bibfield  {author} {\bibinfo {author} {\bibfnamefont {M.}~\bibnamefont
  {Rossmannek}}, \bibinfo {author} {\bibfnamefont {P.~K.}\ \bibnamefont
  {Barkoutsos}}, \bibinfo {author} {\bibfnamefont {P.~J.}\ \bibnamefont
  {Ollitrault}},\ and\ \bibinfo {author} {\bibfnamefont {I.}~\bibnamefont
  {Tavernelli}},\ }\href {https://doi.org/10.1063/5.0029536} {\bibfield
  {journal} {\bibinfo  {journal} {Journal of Chemical Physics}\ }\textbf
  {\bibinfo {volume} {154}},\ \bibinfo {pages} {114105} (\bibinfo {year}
  {2021})},\ \Eprint {https://arxiv.org/abs/2009.01872} {arXiv:2009.01872}
  \BibitemShut {NoStop}%
\bibitem [{\citenamefont {Huggins}\ \emph {et~al.}(2022)\citenamefont
  {Huggins}, \citenamefont {O’Gorman}, \citenamefont {Rubin}, \citenamefont
  {Reichman}, \citenamefont {Babbush},\ and\ \citenamefont
  {Lee}}]{Huggins2022Nature}%
  \BibitemOpen
  \bibfield  {author} {\bibinfo {author} {\bibfnamefont {W.~J.}\ \bibnamefont
  {Huggins}}, \bibinfo {author} {\bibfnamefont {B.~A.}\ \bibnamefont
  {O’Gorman}}, \bibinfo {author} {\bibfnamefont {N.~C.}\ \bibnamefont
  {Rubin}}, \bibinfo {author} {\bibfnamefont {D.~R.}\ \bibnamefont {Reichman}},
  \bibinfo {author} {\bibfnamefont {R.}~\bibnamefont {Babbush}},\ and\ \bibinfo
  {author} {\bibfnamefont {J.}~\bibnamefont {Lee}},\ }\href
  {https://www.nature.com/articles/s41586-021-04351-z} {\bibfield  {journal}
  {\bibinfo  {journal} {Nature}\ }\textbf {\bibinfo {volume} {603}},\ \bibinfo
  {pages} {416} (\bibinfo {year} {2022})}\BibitemShut {NoStop}%
\bibitem [{\citenamefont {Zhang}\ \emph {et~al.}(2022)\citenamefont {Zhang},
  \citenamefont {Huang}, \citenamefont {Sun}, \citenamefont {Lv},\ and\
  \citenamefont {Yuan}}]{YZhang2022arXiv}%
  \BibitemOpen
  \bibfield  {author} {\bibinfo {author} {\bibfnamefont {Y.}~\bibnamefont
  {Zhang}}, \bibinfo {author} {\bibfnamefont {Y.}~\bibnamefont {Huang}},
  \bibinfo {author} {\bibfnamefont {J.}~\bibnamefont {Sun}}, \bibinfo {author}
  {\bibfnamefont {D.}~\bibnamefont {Lv}},\ and\ \bibinfo {author}
  {\bibfnamefont {X.}~\bibnamefont {Yuan}},\ }\href
  {https://arxiv.org/abs/2206.10431} {\bibfield  {journal} {\bibinfo  {journal}
  {arXiv preprint arXiv:2206.10431}\ } (\bibinfo {year} {2022})}\BibitemShut
  {NoStop}%
\bibitem [{\citenamefont {Wan}\ \emph {et~al.}(2023{\natexlab{a}})\citenamefont
  {Wan}, \citenamefont {Huggins}, \citenamefont {Lee},\ and\ \citenamefont
  {Babbush}}]{Wan2023ComMathPhys}%
  \BibitemOpen
  \bibfield  {author} {\bibinfo {author} {\bibfnamefont {K.}~\bibnamefont
  {Wan}}, \bibinfo {author} {\bibfnamefont {W.~J.}\ \bibnamefont {Huggins}},
  \bibinfo {author} {\bibfnamefont {J.}~\bibnamefont {Lee}},\ and\ \bibinfo
  {author} {\bibfnamefont {R.}~\bibnamefont {Babbush}},\ }\href
  {https://arxiv.org/abs/2207.13723} {\bibfield  {journal} {\bibinfo  {journal}
  {Communications in Mathematical Physics}\ ,\ \bibinfo {pages} {1}} (\bibinfo
  {year} {2023}{\natexlab{a}})}\BibitemShut {NoStop}%
\bibitem [{\citenamefont {Verma}\ \emph {et~al.}(2021)\citenamefont {Verma},
  \citenamefont {Huntington}, \citenamefont {Coons}, \citenamefont {Kawashima},
  \citenamefont {Yamazaki},\ and\ \citenamefont {Zaribafiyan}}]{Verma2021JCP}%
  \BibitemOpen
  \bibfield  {author} {\bibinfo {author} {\bibfnamefont {P.}~\bibnamefont
  {Verma}}, \bibinfo {author} {\bibfnamefont {L.}~\bibnamefont {Huntington}},
  \bibinfo {author} {\bibfnamefont {M.~P.}\ \bibnamefont {Coons}}, \bibinfo
  {author} {\bibfnamefont {Y.}~\bibnamefont {Kawashima}}, \bibinfo {author}
  {\bibfnamefont {T.}~\bibnamefont {Yamazaki}},\ and\ \bibinfo {author}
  {\bibfnamefont {A.}~\bibnamefont {Zaribafiyan}},\ }\href
  {https://pubs.aip.org/aip/jcp/article-abstract/155/3/034110/200703/Scaling-up-electronic-structure-calculations-on?redirectedFrom=fulltext}
  {\bibfield  {journal} {\bibinfo  {journal} {The Journal of Chemical Physics}\
  }\textbf {\bibinfo {volume} {155}} (\bibinfo {year} {2021})}\BibitemShut
  {NoStop}%
\bibitem [{\citenamefont {Xu}\ \emph {et~al.}(2023)\citenamefont {Xu},
  \citenamefont {Shimomoto}, \citenamefont {Ten-no},\ and\ \citenamefont
  {Tsuchimochi}}]{EXu2023arXiv}%
  \BibitemOpen
  \bibfield  {author} {\bibinfo {author} {\bibfnamefont {E.}~\bibnamefont
  {Xu}}, \bibinfo {author} {\bibfnamefont {Y.}~\bibnamefont {Shimomoto}},
  \bibinfo {author} {\bibfnamefont {S.~L.}\ \bibnamefont {Ten-no}},\ and\
  \bibinfo {author} {\bibfnamefont {T.}~\bibnamefont {Tsuchimochi}},\ }\href
  {https://arxiv.org/abs/2310.03954} {\bibfield  {journal} {\bibinfo  {journal}
  {arXiv preprint arXiv:2310.03954}\ } (\bibinfo {year} {2023})}\BibitemShut
  {NoStop}%
\bibitem [{\citenamefont {Gonthier}\ \emph {et~al.}(2022)\citenamefont
  {Gonthier}, \citenamefont {Radin}, \citenamefont {Buda}, \citenamefont
  {Doskocil}, \citenamefont {Abuan},\ and\ \citenamefont
  {Romero}}]{Gonthier2022PRR}%
  \BibitemOpen
  \bibfield  {author} {\bibinfo {author} {\bibfnamefont {J.~F.}\ \bibnamefont
  {Gonthier}}, \bibinfo {author} {\bibfnamefont {M.~D.}\ \bibnamefont {Radin}},
  \bibinfo {author} {\bibfnamefont {C.}~\bibnamefont {Buda}}, \bibinfo {author}
  {\bibfnamefont {E.~J.}\ \bibnamefont {Doskocil}}, \bibinfo {author}
  {\bibfnamefont {C.~M.}\ \bibnamefont {Abuan}},\ and\ \bibinfo {author}
  {\bibfnamefont {J.}~\bibnamefont {Romero}},\ }\href
  {https://journals.aps.org/prresearch/abstract/10.1103/PhysRevResearch.4.033154}
  {\bibfield  {journal} {\bibinfo  {journal} {Physical Review Research}\
  }\textbf {\bibinfo {volume} {4}},\ \bibinfo {pages} {033154} (\bibinfo {year}
  {2022})}\BibitemShut {NoStop}%
\bibitem [{\citenamefont {Tilly}\ \emph {et~al.}(2021)\citenamefont {Tilly},
  \citenamefont {Sriluckshmy}, \citenamefont {Patel}, \citenamefont {Fontana},
  \citenamefont {Rungger}, \citenamefont {Grant}, \citenamefont {Anderson},
  \citenamefont {Tennyson},\ and\ \citenamefont {Booth}}]{Tilly2021PRR}%
  \BibitemOpen
  \bibfield  {author} {\bibinfo {author} {\bibfnamefont {J.}~\bibnamefont
  {Tilly}}, \bibinfo {author} {\bibfnamefont {P.}~\bibnamefont {Sriluckshmy}},
  \bibinfo {author} {\bibfnamefont {A.}~\bibnamefont {Patel}}, \bibinfo
  {author} {\bibfnamefont {E.}~\bibnamefont {Fontana}}, \bibinfo {author}
  {\bibfnamefont {I.}~\bibnamefont {Rungger}}, \bibinfo {author} {\bibfnamefont
  {E.}~\bibnamefont {Grant}}, \bibinfo {author} {\bibfnamefont
  {R.}~\bibnamefont {Anderson}}, \bibinfo {author} {\bibfnamefont
  {J.}~\bibnamefont {Tennyson}},\ and\ \bibinfo {author} {\bibfnamefont
  {G.~H.}\ \bibnamefont {Booth}},\ }\href
  {https://journals.aps.org/prresearch/abstract/10.1103/PhysRevResearch.3.033230}
  {\bibfield  {journal} {\bibinfo  {journal} {Physical Review Research}\
  }\textbf {\bibinfo {volume} {3}},\ \bibinfo {pages} {033230} (\bibinfo {year}
  {2021})}\BibitemShut {NoStop}%
\bibitem [{\citenamefont {Takemori}\ \emph {et~al.}(2023)\citenamefont
  {Takemori}, \citenamefont {Teranishi}, \citenamefont {Mizukami},\ and\
  \citenamefont {Yoshioka}}]{takemori2023balancing}%
  \BibitemOpen
  \bibfield  {author} {\bibinfo {author} {\bibfnamefont {N.}~\bibnamefont
  {Takemori}}, \bibinfo {author} {\bibfnamefont {Y.}~\bibnamefont {Teranishi}},
  \bibinfo {author} {\bibfnamefont {W.}~\bibnamefont {Mizukami}},\ and\
  \bibinfo {author} {\bibfnamefont {N.}~\bibnamefont {Yoshioka}},\ }\href@noop
  {} {\bibfield  {journal} {\bibinfo  {journal} {arXiv preprint
  arXiv:2312.17452}\ } (\bibinfo {year} {2023})}\BibitemShut {NoStop}%
\bibitem [{\citenamefont {Kinoshita}\ \emph {et~al.}(2005)\citenamefont
  {Kinoshita}, \citenamefont {Hino},\ and\ \citenamefont
  {Bartlett}}]{Kinoshita2005}%
  \BibitemOpen
  \bibfield  {author} {\bibinfo {author} {\bibfnamefont {T.}~\bibnamefont
  {Kinoshita}}, \bibinfo {author} {\bibfnamefont {O.}~\bibnamefont {Hino}},\
  and\ \bibinfo {author} {\bibfnamefont {R.~J.}\ \bibnamefont {Bartlett}},\
  }\bibfield  {journal} {\bibinfo  {journal} {Journal of Chemical Physics}\
  }\textbf {\bibinfo {volume} {123}},\ \href
  {https://doi.org/10.1063/1.2000251} {10.1063/1.2000251} (\bibinfo {year}
  {2005})\BibitemShut {NoStop}%
\bibitem [{\citenamefont {Hino}\ \emph {et~al.}(2006)\citenamefont {Hino},
  \citenamefont {Kinoshita}, \citenamefont {Chan},\ and\ \citenamefont
  {Bartlett}}]{Hino2006}%
  \BibitemOpen
  \bibfield  {author} {\bibinfo {author} {\bibfnamefont {O.}~\bibnamefont
  {Hino}}, \bibinfo {author} {\bibfnamefont {T.}~\bibnamefont {Kinoshita}},
  \bibinfo {author} {\bibfnamefont {G.~K.-L.}\ \bibnamefont {Chan}},\ and\
  \bibinfo {author} {\bibfnamefont {R.~J.}\ \bibnamefont {Bartlett}},\ }\href
  {https://doi.org/10.1063/1.2180775} {\bibfield  {journal} {\bibinfo
  {journal} {The Journal of Chemical Physics}\ }\textbf {\bibinfo {volume}
  {124}},\ \bibinfo {pages} {114311} (\bibinfo {year} {2006})}\BibitemShut
  {NoStop}%
\bibitem [{\citenamefont {Lyakh}\ \emph {et~al.}(2011)\citenamefont {Lyakh},
  \citenamefont {Lotrich},\ and\ \citenamefont {Bartlett}}]{Lyakh2011CPL}%
  \BibitemOpen
  \bibfield  {author} {\bibinfo {author} {\bibfnamefont {D.~I.}\ \bibnamefont
  {Lyakh}}, \bibinfo {author} {\bibfnamefont {V.~F.}\ \bibnamefont {Lotrich}},\
  and\ \bibinfo {author} {\bibfnamefont {R.~J.}\ \bibnamefont {Bartlett}},\
  }\href
  {https://www.sciencedirect.com/science/article/abs/pii/S0009261410015393}
  {\bibfield  {journal} {\bibinfo  {journal} {Chemical Physics Letters}\
  }\textbf {\bibinfo {volume} {501}},\ \bibinfo {pages} {166} (\bibinfo {year}
  {2011})}\BibitemShut {NoStop}%
\bibitem [{\citenamefont {Melnichuk}\ and\ \citenamefont
  {Bartlett}(2012)}]{Melnichuk2012JCP}%
  \BibitemOpen
  \bibfield  {author} {\bibinfo {author} {\bibfnamefont {A.}~\bibnamefont
  {Melnichuk}}\ and\ \bibinfo {author} {\bibfnamefont {R.~J.}\ \bibnamefont
  {Bartlett}},\ }\href
  {https://pubs.aip.org/aip/jcp/article-abstract/137/21/214103/192246/Relaxed-active-space-Fixing-tailored-CC-with-high?redirectedFrom=fulltext}
  {\bibfield  {journal} {\bibinfo  {journal} {The Journal of chemical physics}\
  }\textbf {\bibinfo {volume} {137}} (\bibinfo {year} {2012})}\BibitemShut
  {NoStop}%
\bibitem [{\citenamefont {Melnichuk}\ and\ \citenamefont
  {Bartlett}(2014)}]{Melnichuk2014JCP}%
  \BibitemOpen
  \bibfield  {author} {\bibinfo {author} {\bibfnamefont {A.}~\bibnamefont
  {Melnichuk}}\ and\ \bibinfo {author} {\bibfnamefont {R.~J.}\ \bibnamefont
  {Bartlett}},\ }\href
  {https://pubs.aip.org/aip/jcp/article-abstract/140/6/064113/193955/Relaxed-active-space-Fixing-tailored-CC-with-high?redirectedFrom=fulltext}
  {\bibfield  {journal} {\bibinfo  {journal} {The Journal of Chemical Physics}\
  }\textbf {\bibinfo {volume} {140}} (\bibinfo {year} {2014})}\BibitemShut
  {NoStop}%
\bibitem [{\citenamefont {M{\"o}rchen}\ \emph {et~al.}(2020)\citenamefont
  {M{\"o}rchen}, \citenamefont {Freitag},\ and\ \citenamefont
  {Reiher}}]{Morchen2020JCP}%
  \BibitemOpen
  \bibfield  {author} {\bibinfo {author} {\bibfnamefont {M.}~\bibnamefont
  {M{\"o}rchen}}, \bibinfo {author} {\bibfnamefont {L.}~\bibnamefont
  {Freitag}},\ and\ \bibinfo {author} {\bibfnamefont {M.}~\bibnamefont
  {Reiher}},\ }\href
  {https://pubs.aip.org/aip/jcp/article-abstract/153/24/244113/200307/Tailored-coupled-cluster-theory-in-varying?redirectedFrom=fulltext}
  {\bibfield  {journal} {\bibinfo  {journal} {The Journal of Chemical Physics}\
  }\textbf {\bibinfo {volume} {153}} (\bibinfo {year} {2020})}\BibitemShut
  {NoStop}%
\bibitem [{\citenamefont {Faulstich}\ \emph {et~al.}(2019)\citenamefont
  {Faulstich}, \citenamefont {M{\'{a}}t{\'{e}}}, \citenamefont {Laestadius},
  \citenamefont {Csirik}, \citenamefont {Veis}, \citenamefont {Antalik},
  \citenamefont {Brabec}, \citenamefont {Schneider}, \citenamefont {Pittner},
  \citenamefont {Kvaal},\ and\ \citenamefont {Legeza}}]{Faulstich2019}%
  \BibitemOpen
  \bibfield  {author} {\bibinfo {author} {\bibfnamefont {F.~M.}\ \bibnamefont
  {Faulstich}}, \bibinfo {author} {\bibfnamefont {M.}~\bibnamefont
  {M{\'{a}}t{\'{e}}}}, \bibinfo {author} {\bibfnamefont {A.}~\bibnamefont
  {Laestadius}}, \bibinfo {author} {\bibfnamefont {M.~A.}\ \bibnamefont
  {Csirik}}, \bibinfo {author} {\bibfnamefont {L.}~\bibnamefont {Veis}},
  \bibinfo {author} {\bibfnamefont {A.}~\bibnamefont {Antalik}}, \bibinfo
  {author} {\bibfnamefont {J.}~\bibnamefont {Brabec}}, \bibinfo {author}
  {\bibfnamefont {R.}~\bibnamefont {Schneider}}, \bibinfo {author}
  {\bibfnamefont {J.}~\bibnamefont {Pittner}}, \bibinfo {author} {\bibfnamefont
  {S.}~\bibnamefont {Kvaal}},\ and\ \bibinfo {author} {\bibfnamefont
  {{\"{O}}.}~\bibnamefont {Legeza}},\ }\href
  {https://pubs.acs.org/doi/full/10.1021/acs.jctc.8b00960} {\bibfield
  {journal} {\bibinfo  {journal} {Journal of Chemical Theory and Computation}\
  }\textbf {\bibinfo {volume} {15}},\ \bibinfo {pages} {2206} (\bibinfo {year}
  {2019})},\ \Eprint {https://arxiv.org/abs/1809.07732} {arXiv:1809.07732}
  \BibitemShut {NoStop}%
\bibitem [{\citenamefont {Vitale}\ \emph {et~al.}(2020)\citenamefont {Vitale},
  \citenamefont {Alavi},\ and\ \citenamefont {Kats}}]{Vitale2020}%
  \BibitemOpen
  \bibfield  {author} {\bibinfo {author} {\bibfnamefont {E.}~\bibnamefont
  {Vitale}}, \bibinfo {author} {\bibfnamefont {A.}~\bibnamefont {Alavi}},\ and\
  \bibinfo {author} {\bibfnamefont {D.}~\bibnamefont {Kats}},\ }\href
  {https://pubs.acs.org/doi/full/10.1021/acs.jctc.0c00470} {\bibfield
  {journal} {\bibinfo  {journal} {Journal of Chemical Theory and Computation}\
  }\textbf {\bibinfo {volume} {16}},\ \bibinfo {pages} {5621} (\bibinfo {year}
  {2020})}\BibitemShut {NoStop}%
\bibitem [{\citenamefont {Vitale}\ \emph {et~al.}(2022)\citenamefont {Vitale},
  \citenamefont {{Li Manni}}, \citenamefont {Alavi},\ and\ \citenamefont
  {Kats}}]{Vitale2022}%
  \BibitemOpen
  \bibfield  {author} {\bibinfo {author} {\bibfnamefont {E.}~\bibnamefont
  {Vitale}}, \bibinfo {author} {\bibfnamefont {G.}~\bibnamefont {{Li Manni}}},
  \bibinfo {author} {\bibfnamefont {A.}~\bibnamefont {Alavi}},\ and\ \bibinfo
  {author} {\bibfnamefont {D.}~\bibnamefont {Kats}},\ }\href
  {https://pubs.acs.org/doi/full/10.1021/acs.jctc.2c00059} {\bibfield
  {journal} {\bibinfo  {journal} {Journal of Chemical Theory and Computation}\
  }\textbf {\bibinfo {volume} {18}},\ \bibinfo {pages} {3427} (\bibinfo {year}
  {2022})}\BibitemShut {NoStop}%
\bibitem [{\citenamefont {Henderson}\ \emph {et~al.}(2014)\citenamefont
  {Henderson}, \citenamefont {Bulik}, \citenamefont {Stein},\ and\
  \citenamefont {Scuseria}}]{Henderson2014JCP}%
  \BibitemOpen
  \bibfield  {author} {\bibinfo {author} {\bibfnamefont {T.~M.}\ \bibnamefont
  {Henderson}}, \bibinfo {author} {\bibfnamefont {I.~W.}\ \bibnamefont
  {Bulik}}, \bibinfo {author} {\bibfnamefont {T.}~\bibnamefont {Stein}},\ and\
  \bibinfo {author} {\bibfnamefont {G.~E.}\ \bibnamefont {Scuseria}},\ }\href
  {https://pubs.aip.org/aip/jcp/article/141/24/244104/915492/Seniority-based-coupled-cluster-theory}
  {\bibfield  {journal} {\bibinfo  {journal} {The Journal of chemical physics}\
  }\textbf {\bibinfo {volume} {141}} (\bibinfo {year} {2014})}\BibitemShut
  {NoStop}%
\bibitem [{\citenamefont {Boguslawski}\ and\ \citenamefont
  {Ayers}(2015)}]{Boguslawski2015linearized}%
  \BibitemOpen
  \bibfield  {author} {\bibinfo {author} {\bibfnamefont {K.}~\bibnamefont
  {Boguslawski}}\ and\ \bibinfo {author} {\bibfnamefont {P.~W.}\ \bibnamefont
  {Ayers}},\ }\href {https://pubs.acs.org/doi/10.1021/acs.jctc.5b00776}
  {\bibfield  {journal} {\bibinfo  {journal} {Journal of Chemical Theory and
  Computation}\ }\textbf {\bibinfo {volume} {11}},\ \bibinfo {pages} {5252}
  (\bibinfo {year} {2015})}\BibitemShut {NoStop}%
\bibitem [{\citenamefont {Leszczyk}\ \emph {et~al.}(2021)\citenamefont
  {Leszczyk}, \citenamefont {M{\'a}t{\'e}}, \citenamefont {Legeza},\ and\
  \citenamefont {Boguslawski}}]{Leszczyk2021JCTC}%
  \BibitemOpen
  \bibfield  {author} {\bibinfo {author} {\bibfnamefont {A.}~\bibnamefont
  {Leszczyk}}, \bibinfo {author} {\bibfnamefont {M.}~\bibnamefont
  {M{\'a}t{\'e}}}, \bibinfo {author} {\bibfnamefont {{\"O}.}~\bibnamefont
  {Legeza}},\ and\ \bibinfo {author} {\bibfnamefont {K.}~\bibnamefont
  {Boguslawski}},\ }\href {https://pubs.acs.org/doi/10.1021/acs.jctc.1c00284}
  {\bibfield  {journal} {\bibinfo  {journal} {Journal of Chemical Theory and
  Computation}\ }\textbf {\bibinfo {volume} {18}},\ \bibinfo {pages} {96}
  (\bibinfo {year} {2021})}\BibitemShut {NoStop}%
\bibitem [{\citenamefont {Nowak}\ \emph {et~al.}(2021)\citenamefont {Nowak},
  \citenamefont {Legeza},\ and\ \citenamefont {Boguslawski}}]{Nowak2021JCP}%
  \BibitemOpen
  \bibfield  {author} {\bibinfo {author} {\bibfnamefont {A.}~\bibnamefont
  {Nowak}}, \bibinfo {author} {\bibfnamefont {{\"O}.}~\bibnamefont {Legeza}},\
  and\ \bibinfo {author} {\bibfnamefont {K.}~\bibnamefont {Boguslawski}},\
  }\href
  {https://pubs.aip.org/aip/jcp/article-abstract/154/8/084111/1023420/Orbital-entanglement-and-correlation-from-pCCD?redirectedFrom=fulltext}
  {\bibfield  {journal} {\bibinfo  {journal} {The Journal of Chemical Physics}\
  }\textbf {\bibinfo {volume} {154}} (\bibinfo {year} {2021})}\BibitemShut
  {NoStop}%
\bibitem [{\citenamefont {Peris}\ \emph
  {et~al.}(1998{\natexlab{a}})\citenamefont {Peris}, \citenamefont
  {Planelles},\ and\ \citenamefont {Paldus}}]{Peris1997IJQC}%
  \BibitemOpen
  \bibfield  {author} {\bibinfo {author} {\bibfnamefont {G.}~\bibnamefont
  {Peris}}, \bibinfo {author} {\bibfnamefont {J.}~\bibnamefont {Planelles}},\
  and\ \bibinfo {author} {\bibfnamefont {J.}~\bibnamefont {Paldus}},\ }\href
  {https://onlinelibrary.wiley.com/doi/abs/10.1002/%28SICI%291097-461X%281997%2962%3A2%3C137%3A%3AAID-QUA2%3E3.0.CO%3B2-X}
  {\bibfield  {journal} {\bibinfo  {journal} {International journal of quantum
  chemistry}\ }\textbf {\bibinfo {volume} {62}},\ \bibinfo {pages} {137}
  (\bibinfo {year} {1998}{\natexlab{a}})}\BibitemShut {NoStop}%
\bibitem [{\citenamefont {Li}\ \emph {et~al.}(1997)\citenamefont {Li},
  \citenamefont {Peris}, \citenamefont {Planelles}, \citenamefont {Rajadall},\
  and\ \citenamefont {Paldus}}]{XLi1997JCP}%
  \BibitemOpen
  \bibfield  {author} {\bibinfo {author} {\bibfnamefont {X.}~\bibnamefont
  {Li}}, \bibinfo {author} {\bibfnamefont {G.}~\bibnamefont {Peris}}, \bibinfo
  {author} {\bibfnamefont {J.}~\bibnamefont {Planelles}}, \bibinfo {author}
  {\bibfnamefont {F.}~\bibnamefont {Rajadall}},\ and\ \bibinfo {author}
  {\bibfnamefont {J.}~\bibnamefont {Paldus}},\ }\href
  {https://pubs.aip.org/aip/jcp/article-abstract/107/1/90/149215/Externally-corrected-singles-and-doubles-coupled?redirectedFrom=fulltext}
  {\bibfield  {journal} {\bibinfo  {journal} {The Journal of chemical physics}\
  }\textbf {\bibinfo {volume} {107}},\ \bibinfo {pages} {90} (\bibinfo {year}
  {1997})}\BibitemShut {NoStop}%
\bibitem [{\citenamefont {Peris}\ \emph
  {et~al.}(1998{\natexlab{b}})\citenamefont {Peris}, \citenamefont {Rajadell},
  \citenamefont {Li}, \citenamefont {Planelles},\ and\ \citenamefont
  {Paldus}}]{Peris1998MolPhys}%
  \BibitemOpen
  \bibfield  {author} {\bibinfo {author} {\bibfnamefont {G.}~\bibnamefont
  {Peris}}, \bibinfo {author} {\bibfnamefont {F.}~\bibnamefont {Rajadell}},
  \bibinfo {author} {\bibfnamefont {X.}~\bibnamefont {Li}}, \bibinfo {author}
  {\bibfnamefont {J.}~\bibnamefont {Planelles}},\ and\ \bibinfo {author}
  {\bibfnamefont {J.}~\bibnamefont {Paldus}},\ }\href
  {https://www.tandfonline.com/doi/abs/10.1080/002689798168529} {\bibfield
  {journal} {\bibinfo  {journal} {Molecular Physics}\ }\textbf {\bibinfo
  {volume} {94}},\ \bibinfo {pages} {235} (\bibinfo {year}
  {1998}{\natexlab{b}})}\BibitemShut {NoStop}%
\bibitem [{\citenamefont {Stolarczyk}(1994)}]{Stolarczyk1994CPL}%
  \BibitemOpen
  \bibfield  {author} {\bibinfo {author} {\bibfnamefont {L.~Z.}\ \bibnamefont
  {Stolarczyk}},\ }\href
  {https://www.sciencedirect.com/science/article/abs/pii/0009261493E1333C}
  {\bibfield  {journal} {\bibinfo  {journal} {Chemical physics letters}\
  }\textbf {\bibinfo {volume} {217}},\ \bibinfo {pages} {1} (\bibinfo {year}
  {1994})}\BibitemShut {NoStop}%
\bibitem [{\citenamefont {Xu}\ and\ \citenamefont {Li}(2015)}]{Xu2015JCP}%
  \BibitemOpen
  \bibfield  {author} {\bibinfo {author} {\bibfnamefont {E.}~\bibnamefont
  {Xu}}\ and\ \bibinfo {author} {\bibfnamefont {S.}~\bibnamefont {Li}},\ }\href
  {https://pubs.aip.org/aip/jcp/article-abstract/142/9/094119/940649/The-externally-corrected-coupled-cluster-approach?redirectedFrom=fulltext}
  {\bibfield  {journal} {\bibinfo  {journal} {The Journal of chemical physics}\
  }\textbf {\bibinfo {volume} {142}} (\bibinfo {year} {2015})}\BibitemShut
  {NoStop}%
\bibitem [{\citenamefont {Deustua}\ \emph {et~al.}(2018)\citenamefont
  {Deustua}, \citenamefont {Magoulas}, \citenamefont {Shen},\ and\
  \citenamefont {Piecuch}}]{Deustua2018JCP}%
  \BibitemOpen
  \bibfield  {author} {\bibinfo {author} {\bibfnamefont {J.~E.}\ \bibnamefont
  {Deustua}}, \bibinfo {author} {\bibfnamefont {I.}~\bibnamefont {Magoulas}},
  \bibinfo {author} {\bibfnamefont {J.}~\bibnamefont {Shen}},\ and\ \bibinfo
  {author} {\bibfnamefont {P.}~\bibnamefont {Piecuch}},\ }\href
  {https://pubs.aip.org/aip/jcp/article/149/15/151101/447932/Communication-Approaching-exact-quantum-chemistry}
  {\bibfield  {journal} {\bibinfo  {journal} {The Journal of Chemical Physics}\
  }\textbf {\bibinfo {volume} {149}} (\bibinfo {year} {2018})}\BibitemShut
  {NoStop}%
\bibitem [{\citenamefont {Magoulas}\ \emph {et~al.}(2021)\citenamefont
  {Magoulas}, \citenamefont {Gururangan}, \citenamefont {Piecuch},
  \citenamefont {Deustua},\ and\ \citenamefont {Shen}}]{Magoulas2021JCTC}%
  \BibitemOpen
  \bibfield  {author} {\bibinfo {author} {\bibfnamefont {I.}~\bibnamefont
  {Magoulas}}, \bibinfo {author} {\bibfnamefont {K.}~\bibnamefont
  {Gururangan}}, \bibinfo {author} {\bibfnamefont {P.}~\bibnamefont {Piecuch}},
  \bibinfo {author} {\bibfnamefont {J.~E.}\ \bibnamefont {Deustua}},\ and\
  \bibinfo {author} {\bibfnamefont {J.}~\bibnamefont {Shen}},\ }\href
  {https://pubs.acs.org/doi/10.1021/acs.jctc.1c00181} {\bibfield  {journal}
  {\bibinfo  {journal} {Journal of Chemical Theory and Computation}\ }\textbf
  {\bibinfo {volume} {17}},\ \bibinfo {pages} {4006} (\bibinfo {year}
  {2021})}\BibitemShut {NoStop}%
\bibitem [{\citenamefont {Lee}\ \emph {et~al.}(2021)\citenamefont {Lee},
  \citenamefont {Zhai}, \citenamefont {Sharma}, \citenamefont {Umrigar},\ and\
  \citenamefont {Chan}}]{SLee2021JCTC}%
  \BibitemOpen
  \bibfield  {author} {\bibinfo {author} {\bibfnamefont {S.}~\bibnamefont
  {Lee}}, \bibinfo {author} {\bibfnamefont {H.}~\bibnamefont {Zhai}}, \bibinfo
  {author} {\bibfnamefont {S.}~\bibnamefont {Sharma}}, \bibinfo {author}
  {\bibfnamefont {C.~J.}\ \bibnamefont {Umrigar}},\ and\ \bibinfo {author}
  {\bibfnamefont {G.~K.-L.}\ \bibnamefont {Chan}},\ }\href
  {https://pubs.acs.org/doi/10.1021/acs.jctc.1c00205} {\bibfield  {journal}
  {\bibinfo  {journal} {Journal of Chemical Theory and Computation}\ }\textbf
  {\bibinfo {volume} {17}},\ \bibinfo {pages} {3414} (\bibinfo {year}
  {2021})}\BibitemShut {NoStop}%
\bibitem [{\citenamefont {Kohda}\ \emph {et~al.}(2022)\citenamefont {Kohda},
  \citenamefont {Imai}, \citenamefont {Kanno}, \citenamefont {Mitarai},
  \citenamefont {Mizukami},\ and\ \citenamefont {Nakagawa}}]{Kohda2022}%
  \BibitemOpen
  \bibfield  {author} {\bibinfo {author} {\bibfnamefont {M.}~\bibnamefont
  {Kohda}}, \bibinfo {author} {\bibfnamefont {R.}~\bibnamefont {Imai}},
  \bibinfo {author} {\bibfnamefont {K.}~\bibnamefont {Kanno}}, \bibinfo
  {author} {\bibfnamefont {K.}~\bibnamefont {Mitarai}}, \bibinfo {author}
  {\bibfnamefont {W.}~\bibnamefont {Mizukami}},\ and\ \bibinfo {author}
  {\bibfnamefont {Y.~O.}\ \bibnamefont {Nakagawa}},\ }\href
  {https://doi.org/10.1103/PhysRevResearch.4.033173} {\bibfield  {journal}
  {\bibinfo  {journal} {Physical Review Research}\ }\textbf {\bibinfo {volume}
  {4}},\ \bibinfo {pages} {033173} (\bibinfo {year} {2022})},\ \Eprint
  {https://arxiv.org/abs/2112.07416} {arXiv:2112.07416} \BibitemShut {NoStop}%
\bibitem [{\citenamefont {Lang}\ \emph {et~al.}(2020)\citenamefont {Lang},
  \citenamefont {Antal{\'{i}}k}, \citenamefont {Veis}, \citenamefont
  {Brandejs}, \citenamefont {Brabec}, \citenamefont {Legeza},\ and\
  \citenamefont {Pittner}}]{Lang2020}%
  \BibitemOpen
  \bibfield  {author} {\bibinfo {author} {\bibfnamefont {J.}~\bibnamefont
  {Lang}}, \bibinfo {author} {\bibfnamefont {A.}~\bibnamefont {Antal{\'{i}}k}},
  \bibinfo {author} {\bibfnamefont {L.}~\bibnamefont {Veis}}, \bibinfo {author}
  {\bibfnamefont {J.}~\bibnamefont {Brandejs}}, \bibinfo {author}
  {\bibfnamefont {J.}~\bibnamefont {Brabec}}, \bibinfo {author} {\bibfnamefont
  {{\"{O}}.}~\bibnamefont {Legeza}},\ and\ \bibinfo {author} {\bibfnamefont
  {J.}~\bibnamefont {Pittner}},\ }\href
  {https://pubs.acs.org/doi/full/10.1021/acs.jctc.0c00065} {\bibfield
  {journal} {\bibinfo  {journal} {Journal of Chemical Theory and Computation}\
  }\textbf {\bibinfo {volume} {16}},\ \bibinfo {pages} {3028} (\bibinfo {year}
  {2020})},\ \Eprint {https://arxiv.org/abs/1907.13466} {arXiv:1907.13466}
  \BibitemShut {NoStop}%
\bibitem [{\citenamefont {Eddins}\ \emph {et~al.}(2022)\citenamefont {Eddins},
  \citenamefont {Motta}, \citenamefont {Gujarati}, \citenamefont {Bravyi},
  \citenamefont {Mezzacapo}, \citenamefont {Hadfield},\ and\ \citenamefont
  {Sheldon}}]{Eddins2022}%
  \BibitemOpen
  \bibfield  {author} {\bibinfo {author} {\bibfnamefont {A.}~\bibnamefont
  {Eddins}}, \bibinfo {author} {\bibfnamefont {M.}~\bibnamefont {Motta}},
  \bibinfo {author} {\bibfnamefont {T.~P.}\ \bibnamefont {Gujarati}}, \bibinfo
  {author} {\bibfnamefont {S.}~\bibnamefont {Bravyi}}, \bibinfo {author}
  {\bibfnamefont {A.}~\bibnamefont {Mezzacapo}}, \bibinfo {author}
  {\bibfnamefont {C.}~\bibnamefont {Hadfield}},\ and\ \bibinfo {author}
  {\bibfnamefont {S.}~\bibnamefont {Sheldon}},\ }\href
  {https://doi.org/10.1103/PRXQuantum.3.010309} {\bibfield  {journal} {\bibinfo
   {journal} {PRX Quantum}\ }\textbf {\bibinfo {volume} {3}},\ \bibinfo {pages}
  {010309} (\bibinfo {year} {2022})}\BibitemShut {NoStop}%
\bibitem [{\citenamefont {Peruzzo}\ \emph {et~al.}(2014)\citenamefont
  {Peruzzo}, \citenamefont {McClean}, \citenamefont {Shadbolt}, \citenamefont
  {Yung}, \citenamefont {Zhou}, \citenamefont {Love}, \citenamefont
  {Aspuru-Guzik},\ and\ \citenamefont {O'Brien}}]{Peruzzo2014}%
  \BibitemOpen
  \bibfield  {author} {\bibinfo {author} {\bibfnamefont {A.}~\bibnamefont
  {Peruzzo}}, \bibinfo {author} {\bibfnamefont {J.}~\bibnamefont {McClean}},
  \bibinfo {author} {\bibfnamefont {P.}~\bibnamefont {Shadbolt}}, \bibinfo
  {author} {\bibfnamefont {M.~H.}\ \bibnamefont {Yung}}, \bibinfo {author}
  {\bibfnamefont {X.~Q.}\ \bibnamefont {Zhou}}, \bibinfo {author}
  {\bibfnamefont {P.~J.}\ \bibnamefont {Love}}, \bibinfo {author}
  {\bibfnamefont {A.}~\bibnamefont {Aspuru-Guzik}},\ and\ \bibinfo {author}
  {\bibfnamefont {J.~L.}\ \bibnamefont {O'Brien}},\ }\href
  {https://doi.org/10.1038/ncomms5213} {\bibfield  {journal} {\bibinfo
  {journal} {Nature Communications 2014 5:1}\ }\textbf {\bibinfo {volume}
  {5}},\ \bibinfo {pages} {1} (\bibinfo {year} {2014})}\BibitemShut {NoStop}%
\bibitem [{\citenamefont {Fedorov}\ \emph {et~al.}(2022)\citenamefont
  {Fedorov}, \citenamefont {Peng}, \citenamefont {Govind},\ and\ \citenamefont
  {Alexeev}}]{Fedorov2022}%
  \BibitemOpen
  \bibfield  {author} {\bibinfo {author} {\bibfnamefont {D.~A.}\ \bibnamefont
  {Fedorov}}, \bibinfo {author} {\bibfnamefont {B.}~\bibnamefont {Peng}},
  \bibinfo {author} {\bibfnamefont {N.}~\bibnamefont {Govind}},\ and\ \bibinfo
  {author} {\bibfnamefont {Y.}~\bibnamefont {Alexeev}},\ }\bibfield  {journal}
  {\bibinfo  {journal} {Materials Theory}\ }\textbf {\bibinfo {volume} {6}},\
  \href {https://doi.org/10.1186/S41313-021-00032-6}
  {10.1186/S41313-021-00032-6} (\bibinfo {year} {2022}),\ \Eprint
  {https://arxiv.org/abs/2103.08505} {arXiv:2103.08505} \BibitemShut {NoStop}%
\bibitem [{\citenamefont {Aspuru-Guzik}\ \emph {et~al.}(2005)\citenamefont
  {Aspuru-Guzik}, \citenamefont {Dutoi}, \citenamefont {Love},\ and\
  \citenamefont {Head-Gordon}}]{Aspuru-Guzik2005}%
  \BibitemOpen
  \bibfield  {author} {\bibinfo {author} {\bibfnamefont {A.}~\bibnamefont
  {Aspuru-Guzik}}, \bibinfo {author} {\bibfnamefont {A.~D.}\ \bibnamefont
  {Dutoi}}, \bibinfo {author} {\bibfnamefont {P.~J.}\ \bibnamefont {Love}},\
  and\ \bibinfo {author} {\bibfnamefont {M.}~\bibnamefont {Head-Gordon}},\
  }\href {https://doi.org/10.1126/SCIENCE.1113479} {\bibfield  {journal}
  {\bibinfo  {journal} {Science}\ }\textbf {\bibinfo {volume} {309}},\ \bibinfo
  {pages} {1704} (\bibinfo {year} {2005})}\BibitemShut {NoStop}%
\bibitem [{\citenamefont {McArdle}\ \emph {et~al.}(2019)\citenamefont
  {McArdle}, \citenamefont {Jones}, \citenamefont {Endo}, \citenamefont {Li},
  \citenamefont {Benjamin},\ and\ \citenamefont {Yuan}}]{McArdle2019}%
  \BibitemOpen
  \bibfield  {author} {\bibinfo {author} {\bibfnamefont {S.}~\bibnamefont
  {McArdle}}, \bibinfo {author} {\bibfnamefont {T.}~\bibnamefont {Jones}},
  \bibinfo {author} {\bibfnamefont {S.}~\bibnamefont {Endo}}, \bibinfo {author}
  {\bibfnamefont {Y.}~\bibnamefont {Li}}, \bibinfo {author} {\bibfnamefont
  {S.~C.}\ \bibnamefont {Benjamin}},\ and\ \bibinfo {author} {\bibfnamefont
  {X.}~\bibnamefont {Yuan}},\ }\href
  {https://doi.org/10.1038/s41534-019-0187-2} {\bibfield  {journal} {\bibinfo
  {journal} {npj Quantum Information 2019 5:1}\ }\textbf {\bibinfo {volume}
  {5}},\ \bibinfo {pages} {1} (\bibinfo {year} {2019})},\ \Eprint
  {https://arxiv.org/abs/1804.03023} {arXiv:1804.03023} \BibitemShut {NoStop}%
\bibitem [{\citenamefont {Motta}\ \emph {et~al.}(2019)\citenamefont {Motta},
  \citenamefont {Sun}, \citenamefont {Tan}, \citenamefont {O'Rourke},
  \citenamefont {Ye}, \citenamefont {Minnich}, \citenamefont {Brand{\~{a}}o},\
  and\ \citenamefont {Chan}}]{Motta2019}%
  \BibitemOpen
  \bibfield  {author} {\bibinfo {author} {\bibfnamefont {M.}~\bibnamefont
  {Motta}}, \bibinfo {author} {\bibfnamefont {C.}~\bibnamefont {Sun}}, \bibinfo
  {author} {\bibfnamefont {A.~T.}\ \bibnamefont {Tan}}, \bibinfo {author}
  {\bibfnamefont {M.~J.}\ \bibnamefont {O'Rourke}}, \bibinfo {author}
  {\bibfnamefont {E.}~\bibnamefont {Ye}}, \bibinfo {author} {\bibfnamefont
  {A.~J.}\ \bibnamefont {Minnich}}, \bibinfo {author} {\bibfnamefont {F.~G.}\
  \bibnamefont {Brand{\~{a}}o}},\ and\ \bibinfo {author} {\bibfnamefont
  {G.~K.~L.}\ \bibnamefont {Chan}},\ }\href
  {https://doi.org/10.1038/s41567-019-0704-4} {\bibfield  {journal} {\bibinfo
  {journal} {Nature Physics 2019 16:2}\ }\textbf {\bibinfo {volume} {16}},\
  \bibinfo {pages} {205} (\bibinfo {year} {2019})},\ \Eprint
  {https://arxiv.org/abs/1901.07653} {arXiv:1901.07653} \BibitemShut {NoStop}%
\bibitem [{\citenamefont {Gomes}\ \emph {et~al.}(2020)\citenamefont {Gomes},
  \citenamefont {Zhang}, \citenamefont {Berthusen}, \citenamefont {Wang},
  \citenamefont {Ho}, \citenamefont {Orth},\ and\ \citenamefont
  {Yao}}]{Gomes2020JCTC}%
  \BibitemOpen
  \bibfield  {author} {\bibinfo {author} {\bibfnamefont {N.}~\bibnamefont
  {Gomes}}, \bibinfo {author} {\bibfnamefont {F.}~\bibnamefont {Zhang}},
  \bibinfo {author} {\bibfnamefont {N.~F.}\ \bibnamefont {Berthusen}}, \bibinfo
  {author} {\bibfnamefont {C.-Z.}\ \bibnamefont {Wang}}, \bibinfo {author}
  {\bibfnamefont {K.-M.}\ \bibnamefont {Ho}}, \bibinfo {author} {\bibfnamefont
  {P.~P.}\ \bibnamefont {Orth}},\ and\ \bibinfo {author} {\bibfnamefont
  {Y.}~\bibnamefont {Yao}},\ }\href
  {https://pubs.acs.org/doi/10.1021/acs.jctc.0c00666} {\bibfield  {journal}
  {\bibinfo  {journal} {Journal of Chemical Theory and Computation}\ }\textbf
  {\bibinfo {volume} {16}},\ \bibinfo {pages} {6256} (\bibinfo {year}
  {2020})}\BibitemShut {NoStop}%
\bibitem [{\citenamefont {Gomes}\ \emph {et~al.}(2021)\citenamefont {Gomes},
  \citenamefont {Mukherjee}, \citenamefont {Zhang}, \citenamefont {Iadecola},
  \citenamefont {Wang}, \citenamefont {Ho}, \citenamefont {Orth},\ and\
  \citenamefont {Yao}}]{Gomes2021AQT}%
  \BibitemOpen
  \bibfield  {author} {\bibinfo {author} {\bibfnamefont {N.}~\bibnamefont
  {Gomes}}, \bibinfo {author} {\bibfnamefont {A.}~\bibnamefont {Mukherjee}},
  \bibinfo {author} {\bibfnamefont {F.}~\bibnamefont {Zhang}}, \bibinfo
  {author} {\bibfnamefont {T.}~\bibnamefont {Iadecola}}, \bibinfo {author}
  {\bibfnamefont {C.-Z.}\ \bibnamefont {Wang}}, \bibinfo {author}
  {\bibfnamefont {K.-M.}\ \bibnamefont {Ho}}, \bibinfo {author} {\bibfnamefont
  {P.~P.}\ \bibnamefont {Orth}},\ and\ \bibinfo {author} {\bibfnamefont
  {Y.-X.}\ \bibnamefont {Yao}},\ }\href
  {https://onlinelibrary.wiley.com/doi/full/10.1002/qute.202100114} {\bibfield
  {journal} {\bibinfo  {journal} {Advanced Quantum Technologies}\ }\textbf
  {\bibinfo {volume} {4}},\ \bibinfo {pages} {2100114} (\bibinfo {year}
  {2021})}\BibitemShut {NoStop}%
\bibitem [{\citenamefont {Tsuchimochi}\ \emph {et~al.}(2023)\citenamefont
  {Tsuchimochi}, \citenamefont {Ryo}, \citenamefont {Ten-No},\ and\
  \citenamefont {Sasasako}}]{Tsuchimochi2023JCTC}%
  \BibitemOpen
  \bibfield  {author} {\bibinfo {author} {\bibfnamefont {T.}~\bibnamefont
  {Tsuchimochi}}, \bibinfo {author} {\bibfnamefont {Y.}~\bibnamefont {Ryo}},
  \bibinfo {author} {\bibfnamefont {S.~L.}\ \bibnamefont {Ten-No}},\ and\
  \bibinfo {author} {\bibfnamefont {K.}~\bibnamefont {Sasasako}},\ }\href
  {https://pubs.acs.org/doi/10.1021/acs.jctc.2c00906} {\bibfield  {journal}
  {\bibinfo  {journal} {Journal of Chemical Theory and Computation}\ }\textbf
  {\bibinfo {volume} {19}},\ \bibinfo {pages} {503} (\bibinfo {year}
  {2023})}\BibitemShut {NoStop}%
\bibitem [{\citenamefont {Izs{\'{a}}k}\ \emph {et~al.}(2023)\citenamefont
  {Izs{\'{a}}k}, \citenamefont {Riplinger}, \citenamefont {Blunt},
  \citenamefont {de~Souza}, \citenamefont {Holzmann}, \citenamefont {Crawford},
  \citenamefont {Camps}, \citenamefont {Neese},\ and\ \citenamefont
  {Schopf}}]{Izsak2023}%
  \BibitemOpen
  \bibfield  {author} {\bibinfo {author} {\bibfnamefont {R.}~\bibnamefont
  {Izs{\'{a}}k}}, \bibinfo {author} {\bibfnamefont {C.}~\bibnamefont
  {Riplinger}}, \bibinfo {author} {\bibfnamefont {N.~S.}\ \bibnamefont
  {Blunt}}, \bibinfo {author} {\bibfnamefont {B.}~\bibnamefont {de~Souza}},
  \bibinfo {author} {\bibfnamefont {N.}~\bibnamefont {Holzmann}}, \bibinfo
  {author} {\bibfnamefont {O.}~\bibnamefont {Crawford}}, \bibinfo {author}
  {\bibfnamefont {J.}~\bibnamefont {Camps}}, \bibinfo {author} {\bibfnamefont
  {F.}~\bibnamefont {Neese}},\ and\ \bibinfo {author} {\bibfnamefont
  {P.}~\bibnamefont {Schopf}},\ }\href {https://doi.org/10.1002/JCC.26958}
  {\bibfield  {journal} {\bibinfo  {journal} {Journal of Computational
  Chemistry}\ }\textbf {\bibinfo {volume} {44}},\ \bibinfo {pages} {406}
  (\bibinfo {year} {2023})},\ \Eprint {https://arxiv.org/abs/2202.04460}
  {arXiv:2202.04460} \BibitemShut {NoStop}%
\bibitem [{che(2023)}]{chemqulacs}%
  \BibitemOpen
  \href@noop {} {\bibinfo {title} {Chemqulacs}},\ \bibinfo {howpublished}
  {\url{https://wmizukami.github.io/chemqulacs/}} (\bibinfo {year}
  {2023})\BibitemShut {NoStop}%
\bibitem [{\citenamefont {Sun}\ \emph {et~al.}(2020)\citenamefont {Sun},
  \citenamefont {Zhang}, \citenamefont {Banerjee}, \citenamefont {Bao},
  \citenamefont {Barbry}, \citenamefont {Blunt}, \citenamefont {Bogdanov},
  \citenamefont {Booth}, \citenamefont {Chen}, \citenamefont {Cui},
  \citenamefont {Eriksen}, \citenamefont {Gao}, \citenamefont {Guo},
  \citenamefont {Hermann}, \citenamefont {Hermes}, \citenamefont {Koh},
  \citenamefont {Koval}, \citenamefont {Lehtola}, \citenamefont {Li},
  \citenamefont {Liu}, \citenamefont {Mardirossian}, \citenamefont {McClain},
  \citenamefont {Motta}, \citenamefont {Mussard}, \citenamefont {Pham},
  \citenamefont {Pulkin}, \citenamefont {Purwanto}, \citenamefont {Robinson},
  \citenamefont {Ronca}, \citenamefont {Sayfutyarova}, \citenamefont
  {Scheurer}, \citenamefont {Schurkus}, \citenamefont {Smith}, \citenamefont
  {Sun}, \citenamefont {Sun}, \citenamefont {Upadhyay}, \citenamefont {Wagner},
  \citenamefont {Wang}, \citenamefont {White}, \citenamefont {Whitfield},
  \citenamefont {Williamson}, \citenamefont {Wouters}, \citenamefont {Yang},
  \citenamefont {Yu}, \citenamefont {Zhu}, \citenamefont {Berkelbach},
  \citenamefont {Sharma}, \citenamefont {Sokolov},\ and\ \citenamefont
  {Chan}}]{Sun2020}%
  \BibitemOpen
  \bibfield  {author} {\bibinfo {author} {\bibfnamefont {Q.}~\bibnamefont
  {Sun}}, \bibinfo {author} {\bibfnamefont {X.}~\bibnamefont {Zhang}}, \bibinfo
  {author} {\bibfnamefont {S.}~\bibnamefont {Banerjee}}, \bibinfo {author}
  {\bibfnamefont {P.}~\bibnamefont {Bao}}, \bibinfo {author} {\bibfnamefont
  {M.}~\bibnamefont {Barbry}}, \bibinfo {author} {\bibfnamefont {N.~S.}\
  \bibnamefont {Blunt}}, \bibinfo {author} {\bibfnamefont {N.~A.}\ \bibnamefont
  {Bogdanov}}, \bibinfo {author} {\bibfnamefont {G.~H.}\ \bibnamefont {Booth}},
  \bibinfo {author} {\bibfnamefont {J.}~\bibnamefont {Chen}}, \bibinfo {author}
  {\bibfnamefont {Z.~H.}\ \bibnamefont {Cui}}, \bibinfo {author} {\bibfnamefont
  {J.~J.}\ \bibnamefont {Eriksen}}, \bibinfo {author} {\bibfnamefont
  {Y.}~\bibnamefont {Gao}}, \bibinfo {author} {\bibfnamefont {S.}~\bibnamefont
  {Guo}}, \bibinfo {author} {\bibfnamefont {J.}~\bibnamefont {Hermann}},
  \bibinfo {author} {\bibfnamefont {M.~R.}\ \bibnamefont {Hermes}}, \bibinfo
  {author} {\bibfnamefont {K.}~\bibnamefont {Koh}}, \bibinfo {author}
  {\bibfnamefont {P.}~\bibnamefont {Koval}}, \bibinfo {author} {\bibfnamefont
  {S.}~\bibnamefont {Lehtola}}, \bibinfo {author} {\bibfnamefont
  {Z.}~\bibnamefont {Li}}, \bibinfo {author} {\bibfnamefont {J.}~\bibnamefont
  {Liu}}, \bibinfo {author} {\bibfnamefont {N.}~\bibnamefont {Mardirossian}},
  \bibinfo {author} {\bibfnamefont {J.~D.}\ \bibnamefont {McClain}}, \bibinfo
  {author} {\bibfnamefont {M.}~\bibnamefont {Motta}}, \bibinfo {author}
  {\bibfnamefont {B.}~\bibnamefont {Mussard}}, \bibinfo {author} {\bibfnamefont
  {H.~Q.}\ \bibnamefont {Pham}}, \bibinfo {author} {\bibfnamefont
  {A.}~\bibnamefont {Pulkin}}, \bibinfo {author} {\bibfnamefont
  {W.}~\bibnamefont {Purwanto}}, \bibinfo {author} {\bibfnamefont {P.~J.}\
  \bibnamefont {Robinson}}, \bibinfo {author} {\bibfnamefont {E.}~\bibnamefont
  {Ronca}}, \bibinfo {author} {\bibfnamefont {E.~R.}\ \bibnamefont
  {Sayfutyarova}}, \bibinfo {author} {\bibfnamefont {M.}~\bibnamefont
  {Scheurer}}, \bibinfo {author} {\bibfnamefont {H.~F.}\ \bibnamefont
  {Schurkus}}, \bibinfo {author} {\bibfnamefont {J.~E.}\ \bibnamefont {Smith}},
  \bibinfo {author} {\bibfnamefont {C.}~\bibnamefont {Sun}}, \bibinfo {author}
  {\bibfnamefont {S.~N.}\ \bibnamefont {Sun}}, \bibinfo {author} {\bibfnamefont
  {S.}~\bibnamefont {Upadhyay}}, \bibinfo {author} {\bibfnamefont {L.~K.}\
  \bibnamefont {Wagner}}, \bibinfo {author} {\bibfnamefont {X.}~\bibnamefont
  {Wang}}, \bibinfo {author} {\bibfnamefont {A.}~\bibnamefont {White}},
  \bibinfo {author} {\bibfnamefont {J.~D.}\ \bibnamefont {Whitfield}}, \bibinfo
  {author} {\bibfnamefont {M.~J.}\ \bibnamefont {Williamson}}, \bibinfo
  {author} {\bibfnamefont {S.}~\bibnamefont {Wouters}}, \bibinfo {author}
  {\bibfnamefont {J.}~\bibnamefont {Yang}}, \bibinfo {author} {\bibfnamefont
  {J.~M.}\ \bibnamefont {Yu}}, \bibinfo {author} {\bibfnamefont
  {T.}~\bibnamefont {Zhu}}, \bibinfo {author} {\bibfnamefont {T.~C.}\
  \bibnamefont {Berkelbach}}, \bibinfo {author} {\bibfnamefont
  {S.}~\bibnamefont {Sharma}}, \bibinfo {author} {\bibfnamefont {A.~Y.}\
  \bibnamefont {Sokolov}},\ and\ \bibinfo {author} {\bibfnamefont {G.~K.~L.}\
  \bibnamefont {Chan}},\ }\href
  {https://pubs.aip.org/aip/jcp/article/153/2/024109/1061482/Recent-developments-in-the-PySCF-program-package}
  {\bibfield  {journal} {\bibinfo  {journal} {Journal of Chemical Physics}\
  }\textbf {\bibinfo {volume} {153}} (\bibinfo {year} {2020})},\ \Eprint
  {https://arxiv.org/abs/2002.12531} {arXiv:2002.12531} \BibitemShut {NoStop}%
\bibitem [{\citenamefont {Neese}(2022)}]{Neese2022}%
  \BibitemOpen
  \bibfield  {author} {\bibinfo {author} {\bibfnamefont {F.}~\bibnamefont
  {Neese}},\ }\href {https://doi.org/10.1002/WCMS.1606} {\bibfield  {journal}
  {\bibinfo  {journal} {Wiley Interdisciplinary Reviews: Computational
  Molecular Science}\ }\textbf {\bibinfo {volume} {12}},\ \bibinfo {pages}
  {e1606} (\bibinfo {year} {2022})}\BibitemShut {NoStop}%
\bibitem [{\citenamefont {Scheurer}\ \emph {et~al.}(2023)\citenamefont
  {Scheurer}, \citenamefont {Anselmetti}, \citenamefont {Oumarou},
  \citenamefont {Gogolin},\ and\ \citenamefont {Rubin}}]{Scheurer2023}%
  \BibitemOpen
  \bibfield  {author} {\bibinfo {author} {\bibfnamefont {M.}~\bibnamefont
  {Scheurer}}, \bibinfo {author} {\bibfnamefont {G.-L.~R.}\ \bibnamefont
  {Anselmetti}}, \bibinfo {author} {\bibfnamefont {O.}~\bibnamefont {Oumarou}},
  \bibinfo {author} {\bibfnamefont {C.}~\bibnamefont {Gogolin}},\ and\ \bibinfo
  {author} {\bibfnamefont {N.~C.}\ \bibnamefont {Rubin}},\ }\href
  {https://arxiv.org/abs/2312.08110v2} {\bibfield  {journal} {\bibinfo
  {journal} {arXiv}\ } (\bibinfo {year} {2023})},\ \Eprint
  {https://arxiv.org/abs/2312.08110} {arXiv:2312.08110} \BibitemShut {NoStop}%
\bibitem [{\citenamefont {Ventura}\ \emph {et~al.}(2003)\citenamefont
  {Ventura}, \citenamefont {{Do Monte}}, \citenamefont {Dallos},\ and\
  \citenamefont {Lischka}}]{Ventura2003}%
  \BibitemOpen
  \bibfield  {author} {\bibinfo {author} {\bibfnamefont {E.}~\bibnamefont
  {Ventura}}, \bibinfo {author} {\bibfnamefont {S.~A.}\ \bibnamefont {{Do
  Monte}}}, \bibinfo {author} {\bibfnamefont {M.}~\bibnamefont {Dallos}},\ and\
  \bibinfo {author} {\bibfnamefont {H.}~\bibnamefont {Lischka}},\ }\href
  {https://pubs.acs.org/doi/full/10.1021/jp0259014} {\bibfield  {journal}
  {\bibinfo  {journal} {Journal of Physical Chemistry A}\ }\textbf {\bibinfo
  {volume} {107}},\ \bibinfo {pages} {1175} (\bibinfo {year}
  {2003})}\BibitemShut {NoStop}%
\bibitem [{\citenamefont {Shepard}\ \emph {et~al.}(1992)\citenamefont
  {Shepard}, \citenamefont {Lischka}, \citenamefont {Szalay}, \citenamefont
  {Kovar},\ and\ \citenamefont {Ernzerhof}}]{Shepard1991}%
  \BibitemOpen
  \bibfield  {author} {\bibinfo {author} {\bibfnamefont {R.}~\bibnamefont
  {Shepard}}, \bibinfo {author} {\bibfnamefont {H.}~\bibnamefont {Lischka}},
  \bibinfo {author} {\bibfnamefont {P.}~\bibnamefont {Szalay}}, \bibinfo
  {author} {\bibfnamefont {T.}~\bibnamefont {Kovar}},\ and\ \bibinfo {author}
  {\bibfnamefont {M.}~\bibnamefont {Ernzerhof}},\ }\href
  {https://doi.org/10.1063/1.462060} {\bibfield  {journal} {\bibinfo  {journal}
  {J. Chem. Phys.}\ }\textbf {\bibinfo {volume} {96}},\ \bibinfo {pages} {2085}
  (\bibinfo {year} {1992})}\BibitemShut {NoStop}%
\bibitem [{\citenamefont {Staroverov}\ and\ \citenamefont
  {Davidson}(2001)}]{Staroverov2001}%
  \BibitemOpen
  \bibfield  {author} {\bibinfo {author} {\bibfnamefont {V.~N.}\ \bibnamefont
  {Staroverov}}\ and\ \bibinfo {author} {\bibfnamefont {E.~R.}\ \bibnamefont
  {Davidson}},\ }\href {https://doi.org/10.1016/S0166-1280(01)00536-X}
  {\bibfield  {journal} {\bibinfo  {journal} {Journal of Molecular Structure:
  THEOCHEM}\ }\textbf {\bibinfo {volume} {573}},\ \bibinfo {pages} {81}
  (\bibinfo {year} {2001})}\BibitemShut {NoStop}%
\bibitem [{\citenamefont {{von E. Doering}}\ \emph {et~al.}(1971)\citenamefont
  {{von E. Doering}}, \citenamefont {Toscano},\ and\ \citenamefont
  {Beasley}}]{Von1971}%
  \BibitemOpen
  \bibfield  {author} {\bibinfo {author} {\bibfnamefont {W.}~\bibnamefont {{von
  E. Doering}}}, \bibinfo {author} {\bibfnamefont {V.}~\bibnamefont
  {Toscano}},\ and\ \bibinfo {author} {\bibfnamefont {G.}~\bibnamefont
  {Beasley}},\ }\href {https://doi.org/10.1016/S0040-4020(01)91694-1}
  {\bibfield  {journal} {\bibinfo  {journal} {Tetrahedron}\ }\textbf {\bibinfo
  {volume} {27}},\ \bibinfo {pages} {5299} (\bibinfo {year}
  {1971})}\BibitemShut {NoStop}%
\bibitem [{\citenamefont {Kanno}\ \emph {et~al.}(2023)\citenamefont {Kanno},
  \citenamefont {Kohda}, \citenamefont {Imai}, \citenamefont {Koh},
  \citenamefont {Mitarai}, \citenamefont {Mizukami},\ and\ \citenamefont
  {Nakagawa}}]{kanno2023quantum}%
  \BibitemOpen
  \bibfield  {author} {\bibinfo {author} {\bibfnamefont {K.}~\bibnamefont
  {Kanno}}, \bibinfo {author} {\bibfnamefont {M.}~\bibnamefont {Kohda}},
  \bibinfo {author} {\bibfnamefont {R.}~\bibnamefont {Imai}}, \bibinfo {author}
  {\bibfnamefont {S.}~\bibnamefont {Koh}}, \bibinfo {author} {\bibfnamefont
  {K.}~\bibnamefont {Mitarai}}, \bibinfo {author} {\bibfnamefont
  {W.}~\bibnamefont {Mizukami}},\ and\ \bibinfo {author} {\bibfnamefont
  {Y.~O.}\ \bibnamefont {Nakagawa}},\ }\href {https://arxiv.org/abs/2302.11320}
  {\bibfield  {journal} {\bibinfo  {journal} {arXiv preprint arXiv:2302.11320}\
  } (\bibinfo {year} {2023})}\BibitemShut {NoStop}%
\bibitem [{\citenamefont {Nakagawa}\ \emph {et~al.}(2023)\citenamefont
  {Nakagawa}, \citenamefont {Kamoshita}, \citenamefont {Mizukami},
  \citenamefont {Sudo},\ and\ \citenamefont {Ohnishi}}]{Nakagawa2023arXiv}%
  \BibitemOpen
  \bibfield  {author} {\bibinfo {author} {\bibfnamefont {Y.~O.}\ \bibnamefont
  {Nakagawa}}, \bibinfo {author} {\bibfnamefont {M.}~\bibnamefont {Kamoshita}},
  \bibinfo {author} {\bibfnamefont {W.}~\bibnamefont {Mizukami}}, \bibinfo
  {author} {\bibfnamefont {S.}~\bibnamefont {Sudo}},\ and\ \bibinfo {author}
  {\bibfnamefont {Y.-y.}\ \bibnamefont {Ohnishi}},\ }\href
  {https://arxiv.org/abs/2311.01105} {\bibfield  {journal} {\bibinfo  {journal}
  {arXiv preprint arXiv:2311.01105}\ } (\bibinfo {year} {2023})}\BibitemShut
  {NoStop}%
\bibitem [{\citenamefont {Liao}\ \emph {et~al.}(2024)\citenamefont {Liao},
  \citenamefont {Ding},\ and\ \citenamefont {Schilling}}]{liao2024unveiling}%
  \BibitemOpen
  \bibfield  {author} {\bibinfo {author} {\bibfnamefont {K.}~\bibnamefont
  {Liao}}, \bibinfo {author} {\bibfnamefont {L.}~\bibnamefont {Ding}},\ and\
  \bibinfo {author} {\bibfnamefont {C.}~\bibnamefont {Schilling}},\ }\href@noop
  {} {\bibfield  {journal} {\bibinfo  {journal} {arXiv preprint
  arXiv:2402.16841}\ } (\bibinfo {year} {2024})}\BibitemShut {NoStop}%
\bibitem [{\citenamefont {Feldmann}\ \emph {et~al.}(2024)\citenamefont
  {Feldmann}, \citenamefont {Mörchen}, \citenamefont {Lang}, \citenamefont
  {Lesiuk},\ and\ \citenamefont {Reiher}}]{feldmann2024complete}%
  \BibitemOpen
  \bibfield  {author} {\bibinfo {author} {\bibfnamefont {R.}~\bibnamefont
  {Feldmann}}, \bibinfo {author} {\bibfnamefont {M.}~\bibnamefont {Mörchen}},
  \bibinfo {author} {\bibfnamefont {J.}~\bibnamefont {Lang}}, \bibinfo {author}
  {\bibfnamefont {M.}~\bibnamefont {Lesiuk}},\ and\ \bibinfo {author}
  {\bibfnamefont {M.}~\bibnamefont {Reiher}},\ }\href@noop {} {\bibfield
  {journal} {\bibinfo  {journal} {arXiv preprint arXiv:2404.06070}\ } (\bibinfo
  {year} {2024})}\BibitemShut {NoStop}%
\bibitem [{\citenamefont {Zhao}\ \emph {et~al.}(2021)\citenamefont {Zhao},
  \citenamefont {Rubin},\ and\ \citenamefont {Miyake}}]{zhao2021fermionic}%
  \BibitemOpen
  \bibfield  {author} {\bibinfo {author} {\bibfnamefont {A.}~\bibnamefont
  {Zhao}}, \bibinfo {author} {\bibfnamefont {N.~C.}\ \bibnamefont {Rubin}},\
  and\ \bibinfo {author} {\bibfnamefont {A.}~\bibnamefont {Miyake}},\
  }\href@noop {} {\bibfield  {journal} {\bibinfo  {journal} {Physical Review
  Letters}\ }\textbf {\bibinfo {volume} {127}},\ \bibinfo {pages} {110504}
  (\bibinfo {year} {2021})}\BibitemShut {NoStop}%
\bibitem [{\citenamefont {Wan}\ \emph {et~al.}(2023{\natexlab{b}})\citenamefont
  {Wan}, \citenamefont {Huggins}, \citenamefont {Lee},\ and\ \citenamefont
  {Babbush}}]{wan2023matchgate}%
  \BibitemOpen
  \bibfield  {author} {\bibinfo {author} {\bibfnamefont {K.}~\bibnamefont
  {Wan}}, \bibinfo {author} {\bibfnamefont {W.~J.}\ \bibnamefont {Huggins}},
  \bibinfo {author} {\bibfnamefont {J.}~\bibnamefont {Lee}},\ and\ \bibinfo
  {author} {\bibfnamefont {R.}~\bibnamefont {Babbush}},\ }\href@noop {}
  {\bibfield  {journal} {\bibinfo  {journal} {Communications in Mathematical
  Physics}\ }\textbf {\bibinfo {volume} {404}},\ \bibinfo {pages} {629}
  (\bibinfo {year} {2023}{\natexlab{b}})}\BibitemShut {NoStop}%
\end{thebibliography}%

\end{document}